\documentclass[opre,nonblindrev]{informs3}

\OneAndAHalfSpacedXII

\usepackage{endnotes}
\let\enotesize=\normalsize
\def\notesname{Endnotes}%
\def\makeenmark{\hbox to1.275em{\theenmark.\enskip\hss}}
\def\enoteformat{\rightskip0pt\leftskip0pt\parindent=1.275em
  \leavevmode\llap{\makeenmark}}

\usepackage{natbib}
 \bibpunct[, ]{(}{)}{,}{a}{}{,}%
 \def\bibfont{\small}%
\TheoremsNumberedThrough     
\ECRepeatTheorems

\MANUSCRIPTNO{}

\renewcommand{\theARTICLETOP}{}
\renewcommand{\theRRHFirstLine}{}
\renewcommand{\theRRHSecondLine}{}
\renewcommand{\theLRHFirstLine}{}
\renewcommand{\theLRHSecondLine}{}

\EquationsNumberedThrough    
\usepackage{amsfonts,latexsym,xspace,makeidx, amssymb,bm}
\usepackage{graphicx}
\usepackage{subfigure}
\usepackage{epstopdf}
\usepackage{algorithmic,algorithm}
\usepackage[hidelinks]{hyperref}
\usepackage{booktabs}
\usepackage{multirow}
\usepackage{mathtools}
\usepackage{color,cases}
\usepackage{boxedminipage}
\usepackage{fp,fp-exp}
\usepackage{subfigure}
\usepackage{nicefrac}
\usepackage{marvosym}

\renewcommand\qed{\BoldSquare}

\newcommand\E{\mathbb{E}}
\newcommand\R{\mathbb{R}}

\newcommand\poly{\operatorname{poly}}

\newcommand{\Exp}{\mathop{\mathbb E}\displaylimits}
\newcommand{\Var}{\mathop{\mathbb V}\displaylimits}

\newcommand{\Ed}{\bar{d}}
\newcommand{\Es}{\bar{s}}

\newcommand{\barn}{\bar{n}}

\def\shownotes{1}  
\ifnum\shownotes=1
\newcommand{\authnote}[2]{{$\ll$\textsf{#1 notes: #2}$\gg$}}
\else
\newcommand{\authnote}[2]{}
\fi
\usepackage{color}
\definecolor{lawngreen}{RGB}{0,250,154}

\FPeval{\ratio}{round(1/10,1)}
\FPeval{\constT}{round(2,0)}

\newcommand{\cp}{5}
\FPeval{\cpv}{5}

\FPeval{\cptwov}{10}

\FPeval{\deltav}{3}

\newcommand{\cth}{c_1}
\FPeval{\cthv}{4}

\newcommand{\cb}{c_{7}}
\FPeval{\cbv}{round(1/6, 2)}
\FPeval{\cbv}{round((1-e^(-\cthv))/\cthv, 2)}

\FPeval{\cdv}{round(\cbv/\deltav/10, 5)}

\newcommand{\ckdisp}{5/6}
\FPeval{\ckappav}{round(\cp/(1+\cp),3)}

\FPeval{\cdpv}{round(\cbv/\deltav/12*\ckappav,5)}

\FPeval{\cev}{round(\cdv/2, 6)}

\newcommand{\cea}{c_{2}}
\FPeval{\ceav}{round(\cev/\cpv, 7)}

\newcommand{\cf}{c_{3}}
\FPeval{\cfv}{round(-(\ceav + 5/216*(0.894839316814371-1)), 5)}

\newcommand{\cmx}{c}
\FPeval{\cmxv}{round(1/\deltav/\deltav/36/4, 5)}

\newcommand{\ck}{c_{\mathcal{L}}}
\newcommand{\ch}{c_{4}}
\FPeval{\tempaa}{round(1/3- e^(-\cthv),4)}
\FPeval{\tempab}{round(1/\deltav/3- 2*e^(-\cthv),4)}
\FPeval{\chv}{round(\tempaa^2/18 -.00001,5)}
\newcommand{\cha}{c_{6}}
\FPeval{\chav}{round(\tempaa^2/12-.00001,5)}

\newcommand{\cfh}{c_{5}}
\FPeval{\cfhv}{min(\cfv,\chv)}
\FPeval{\tb}{round(6/5*\cpv,0)}

\FPeval{\tbb}{round(6/5*\cpv+1,0)}
\newcommand{\cgamma}{c_0}
\FPeval{\cgammav}{round(max(max(\tbb/\cfhv,\tbb/\chav), 2/\ceav) + 1,0)}

\FPeval{\cqv}{3}

\newcommand {\calG}   {{\cal{G}}}

\newcommand {\calZ}   {{\cal{Z}}}
\newcommand{\eps}{\epsilon}

\usepackage{prettyref}

\newcommand{\savehyperref}[2]{\texorpdfstring{\hyperref[#1]{#2}}{#2}}

\newrefformat{eq}{\savehyperref{#1}{\textup{(\ref*{#1})}}}
\newrefformat{lem}{\savehyperref{#1}{Lemma~\ref*{#1}}}
\newrefformat{lemma}{\savehyperref{#1}{Lemma~\ref*{#1}}}
\newrefformat{def}{\savehyperref{#1}{Definition~\ref*{#1}}}
\newrefformat{thm}{\savehyperref{#1}{Theorem~\ref*{#1}}}
\newrefformat{cor}{\savehyperref{#1}{Corollary~\ref*{#1}}}
\newrefformat{corr}{\savehyperref{#1}{Corollary~\ref*{#1}}}
\newrefformat{cha}{\savehyperref{#1}{Chapter~\ref*{#1}}}
\newrefformat{sec}{\savehyperref{#1}{Section~\ref*{#1}}}
\newrefformat{app}{\savehyperref{#1}{Appendix~\ref*{#1}}}
\newrefformat{tab}{\savehyperref{#1}{Table~\ref*{#1}}}
\newrefformat{fig}{\savehyperref{#1}{Figure~\ref*{#1}}}
\newrefformat{hyp}{\savehyperref{#1}{Hypothesis~\ref*{#1}}}
\newrefformat{alg}{\savehyperref{#1}{Algorithm~\ref*{#1}}}
\newrefformat{item}{\savehyperref{#1}{Item~\ref*{#1}}}
\newrefformat{step}{\savehyperref{#1}{step~\ref*{#1}}}
\newrefformat{conj}{\savehyperref{#1}{Conjecture~\ref*{#1}}}
\newrefformat{fact}{\savehyperref{#1}{Fact~\ref*{#1}}}
\newrefformat{prop}{\savehyperref{#1}{Proposition~\ref*{#1}}}
\newrefformat{proposition}{\savehyperref{#1}{Proposition~\ref*{#1}}}
\newrefformat{claim}{\savehyperref{#1}{Claim~\ref*{#1}}}
\newrefformat{example}{\savehyperref{#1}{Example~\ref*{#1}}}
\newrefformat{assumption}{\savehyperref{#1}{Assumption~\ref*{#1}}}

\let\pref=\prettyref

\newcommand{\Sref}[1]{\hyperref[#1]{Section \ref*{#1}}}

\renewcommand{\qed}{\hfill \ensuremath{\Box}}

\begin{document}


\RUNAUTHOR{Chen, Ma, Zhang and Zhou}

\RUNTITLE{Optimal Design of Process Flexibility for General Production Systems}

\TITLE{Optimal Design of Process Flexibility for General Production Systems}

\ARTICLEAUTHORS{
\AUTHOR{Xi Chen}
\AFF{New York University, New York, NY, 10012, \EMAIL{xchen3@stern.nyu.edu}}
\AUTHOR{Tengyu Ma}
\AFF{Facebook AI Research, Menlo Park, CA, 94025, \EMAIL{tengyuma@stanford.edu}}
\AUTHOR{Jiawei Zhang}
\AFF{New York University, New York, NY, 10012\\
NYU Shanghai, Shanghai, China 200122 \EMAIL{jzhang@stern.nyu.edu}}
\AUTHOR{Yuan Zhou}
\AFF{Indiana University at Bloomington, Bloomington, IN, 47405\\ University of Illinois Urbana-Champaign, Urbana, IL, 61801, \EMAIL{yzhoucs@indiana.edu}}
}

\ABSTRACT{Process flexibility is widely adopted as an effective strategy for responding to uncertain demand. Many algorithms for constructing sparse flexibility designs with good theoretical guarantees have been developed for balanced and symmetrical production systems. These systems assume that the number of plants equals the number of products, that supplies have the same capacity, and that demands are independently and identically distributed.

In this paper, we relax these assumptions and consider a general class of production systems.  We construct a simple flexibility design to fulfill $(1-\epsilon)$-fraction of expected demand with high probability (w.h.p.) where the average degree is $O(\ln(1/\eps))$. To motivate our construction, we first consider a natural weighted probabilistic construction from \cite{Chou:11} where the degree of each node is proportional to its expected capacity.
However, this strategy is shown to be sub-optimal.  To obtain an optimal construction, we develop a simple yet effective thresholding scheme.   The analysis of our approach extends the  classical analysis of expander graphs by overcoming several technical difficulties. Our approach may prove useful in other applications that require expansion properties of graphs with non-uniform degree sequences.
}

\maketitle

\section{Introduction}

\begin{figure}[!t]
\caption{Fulfilled demand $Z_G(s,d)$ as the value of maximum flow from the supply side to the demand side.}
        \centering
     \includegraphics[width=0.57\textwidth]{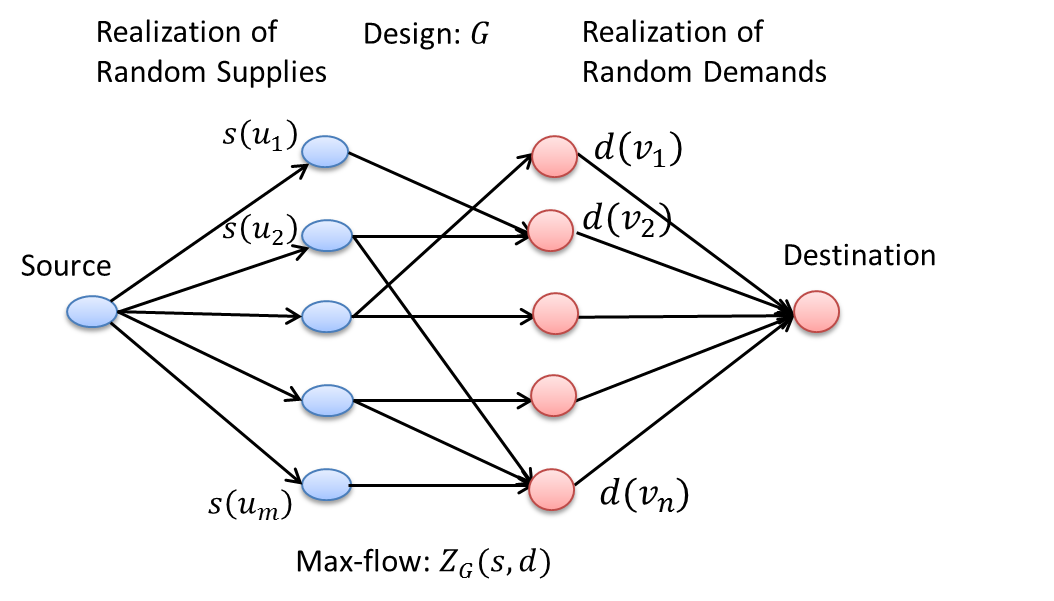}
        \label{fig:max-flow}
        \vspace{-2mm}
\end{figure}

Process flexibility (a.k.a., capacity pooling) is a successful operational strategy in manufacturing industries to hedge against demand uncertainty. The classical work of \cite{Jordan:95} models a manufacturing process flexibility design as a bipartite graph $G=(U\cup V, E)$, where $U$ denotes a set of $m$ supply nodes (representing production plants) and $V$ denotes the set of $n$ demand nodes (representing products demanded in the market). Edge $(u,v)\in E$ appears if node $u$ can supply node $v$ (or plant $u$ can produce product $v$). Node $u \in U$   has a supply (i.e., production capability) characterized by the random variable $s(u)$ and node $v \in V$ generates a random demand $d(v)$.%
\footnote{Supply uncertainty is common in practice. For example, the random failure or shutdown will cause the supply to be random.}
{ The task of designing process flexibility concerns about the construction of the edge set $E$ in $G$.}
Given a flexibility design $G$, and realizations of supplies $\{s(u)\}_{u \in U}$ and demands $\{d(v)\}_{v \in V}$, the total fulfilled demand $\mathcal{Z}_G(s, d)$ is the value of the maximum flow from supply to demand nodes (see \pref{fig:max-flow}).   A production system has full flexibility if any plant can produce all of the products (i.e., $G$ is a complete bipartite graph). Although full flexibility can best fulfill uncertain demand, { it comes at the expense of drastically increased implementation and/or operational costs}. An ideal design strikes the right balance between  these costs and ability to meet uncertain demand. Suppliers typically prefer sparse designs, where each plant is only capable of producing a small number of properly chosen products. In this paper, we aim to \emph{construct the { (asymptotically)} sparsest design $G$ such that the fulfilled demand  can almost match the total expected demand with high probability (w.h.p.).}

Due to the importance of this problem in manufacturing, a number of papers tackle this question by constructing sparse designs and analyzing their performances (see, e.g., \cite{Chou:10}, \cite{Chou:11}, \cite{Simchi:12a}, \cite{Simchi:12b}, \cite{Deng:13}, \cite{Wang:13}, \cite{Chen:15}, \cite{Shi15Multi}, \cite{Tsitsiklis:15}, \cite{Bidkhori:16}, \cite{Zhang:16:long}, and references therein). Most papers study the special setting where (1) the production system\footnote{We use \emph{production system} to denote the set of supply and demand nodes and their associated supply and demand random variables.} is \emph{balanced} (the number of supply and demand nodes are equal, i.e., $m=n$) and (2) \emph{symmetrical} (the random supplies $s(u)$ share the same deterministic capacity or are identically independently distributed (\emph{i.i.d.}) and demands $d(v)$ are \emph{i.i.d.}. These assumptions facilitate the construction of designs  and corresponding theoretical analysis (e.g., popular chaining designs for balanced systems) at the cost of realism. The number of products is typically orders of magnitude larger than the number of plants. Plants typically differ in capacity and demands for products can also differ significantly. Some recent papers tackle production systems that are unbalanced and asymmetrical. { For example, when the ratio between each realized demand and its expectation is constant, the work by \cite{Chou:10}  proposes a  probabilistic construction that achieves $(1-\epsilon)$-optimality (w.r.t. the full flexibility) in expectation with the average degree of $O(1/\eps)$. \cite{Chou:11} further develops a \emph{weighted probabilistic  construction} { (see the proof of Theorem 5)} that links a supply node with a demand node with a probability proportional to their expected capacities\footnote{For the ease of presentation, expected capacity refers to either expected supply or expected demand when the context is clear.}. This construction with average degree $O(1/\eps)$ achieves $(1-\epsilon)$-optimality for all demand scenarios.} \label{page:chou}

\cite{Deng:13} provided several design guidelines for unbalanced but symmetrical systems based on extensive simulation studies.  \cite{Bidkhori:16} derived a distribution-free lower bound for a generalized chaining structure using the mean and partial expectation of the demand. \cite{Shi15Multi} considered a multi-period general production system and proposed the ``generalized chaining condition" to measure the effectiveness of a flexibility design. However, the multi-period setting in \cite{Shi15Multi} assumes that the unsatisfied demands are \emph{backlogged} and is thus fundamentally different from our problem where unsatisfied demand is lost.  \cite{Tsitsiklis:15}
studied the flexibility design problem for a multi-server queuing model and showed that, with limited flexibility, it is possible to simultaneously achieve a large capacity region and an asymptotically vanishing delay.

\subsection{Research Goal}
\label{sec:goal}
In this paper, we focus on the construction of an optimal flexibility design in a single-period setting. In particular, for a general system, our goal is to construct a sparse flexibility design $G$ that is  $(1-\epsilon)$-optimal with high probability (w.h.p.).

More specifically, let us denote the expectation of the random demand of $v \in V$ and the random supply $u \in U$ by $\bar{d}(v)$ and $\bar{s}(u)$, respectively; that is,  $\bar{d}(v)=\E d(v)$ and $\bar{s}(u)=\E s(u)$.  We assume  the  expected total supply matches the expected total demand, i.e.,  $\sum_{u \in U} \bar{s}(u)= \sum_{v \in V} \bar{d}(v)$. As the expected total supply and demand are the same,  we normalize them to be one by dividing each $\bar{s}(u)$ by $\sum \bar{s}(u)$ and each $\bar{d}(v)$ by $\sum \bar{d}(v)$. That is, $\sum_{u \in U}\Es(u) = \sum_{v \in V} \Ed(v) =1$. This normalization step does not result in a loss of generality. As the maximum flow is linear in the supply and demand vectors, the normalization step does not change the ratio between maximum flow under a sparse flexibility design and the full flexibility design. A design $G$ is said to be $(1-\epsilon)$-optimal w.h.p. when the fulfilled demand matches at least $(1-\epsilon)$-fraction of the expected total demand w.h.p. That is,
\begin{equation}
  \Pr_{s(\cdot), d(\cdot)}\left[\calZ_G(s,d) \geq 1-\eps  \right]  \geq 1- \zeta,
  \label{eq:goal}
\end{equation}
for some small $\epsilon>0$ and $\zeta=O(\bar{n}^{-C})$ with some universal constant $C>0$.\footnote{We note that the term \emph{w.h.p.} requires that the event considered (e.g., $\calZ_G(s,d) \geq 1-\eps$) holds not only a with probability tending to $1$, but also at the rate of $1-  \bar{n}^{-\Omega(1)}$ (see the definition of ``w.h.p." in Definition of 1.1.2. in \cite{Tao12RM} and Section \ref{sec:org} for the asymptotic notations $O(\cdot)$, $\Omega(\cdot)$, $\Theta(\cdot)$, $o(\cdot)$, and $\omega(\cdot)$).}  Here, we define $\bar{n}=\max(m,n)$.  The randomness  in \eqref{eq:goal} comes from the random supply vector $s(\cdot) \triangleq \{s(u)\}_{u\in U}$ and demand vector $d(\cdot) \triangleq \{d(v)\}_{v \in V}$.

{ Our research goal is for a wide class of production systems (characterized by Assumption \ref{enum:assumption-boundedness} below in Section \ref{sec:assump}), and for any optimality parameter $\epsilon>0$, to construct a design $G = (U \cup V, E)$ with an as small as possible average degree that is  $(1-\epsilon)$-optimal w.h.p.}




\subsection{Main Results -- Optimal Construction and Technical Contributions}

\cite{Chou:11} considered a weighted probabilistic construction (WPC) that links a pair of nodes $(u,v)$ with probability $r(u,v) \propto \bar{s}(u) \bar{d}(v)$. Here, the symbol ``$\propto$''  means that $r(u,v)/ (\bar{s}(u)\bar{d}(v))$ is a constant for any pair of $(u,v)$. The idea behind this construction is intuitive: a pair of nodes $(u,v)$ with either a  large expected supply $\Es(u)$ or a large expected demand $\Ed(v)$ (or both) should have a higher probability of being linked.
In a balanced and symmetrical system, the WPC naturally reduces to a uniform probabilistic construction. By choosing the linkage probability so that each node has an average degree of $O(\ln(1/\epsilon))$,  \cite{Chen:15} showed that the uniform probabilistic construction achieves $(1-\epsilon)$-optimality w.h.p. Moreover, such a construction is asymptotically optimal, which means it has the fewest possible edges for achieving $(1-\epsilon)$-optimality w.h.p. up to a constant factor. Motivated by this success story for symmetrical and balanced systems, a question naturally  arises: for a general system, can the WPC still achieve  $(1-\epsilon)$-optimality w.h.p. with an average degree of $O(\ln(1/\epsilon))$?

However, this question has a negative answer. When using the WPC, the expected degree of a node $u \in U$ is proportional to $\Es(u)$ and that of a node $v \in V$ is proportional to $\Ed(v)$.  In an asymmetrical system with heterogeneous expected supplies and demands, if $\Es(u)$ or $\Ed(v)$ are small, node $u$ or $v$ may be isolated under the WPC. If so, the supply of those isolated $u$ or demand at those isolated $v$ can never be fulfilled. With this intuition in mind, in \pref{thm:WPC-suboptimal} and \pref{sec:PPC}, we construct an instance where the WPC with the average degree $O(\ln(1/\epsilon))$ leads to the isolation of many nodes with small expected capacities, and thus fails to achieve $(1-\epsilon)$-optimality w.h.p. For this, the WPC requires the average degree to be at least $\Omega(1/\eps)$ to achieve $(1-\eps)$-optimality w.h.p.

A natural fix for the WPC would be forcing every node to have a degree of at least one. This simple fix removes isolated nodes while the degrees are still ``roughly proportional'' to the node mean capacities. However, as we argue in the final paragraph of \pref{sec:PPC} below, in some cases, nodes with small capacities (``small nodes" for short) may need degrees almost as high as those of nodes with large mean capacities (``large nodes'').

In sum, we realize the following important facts, which make constructing an optimal flexibility design for a general system fundamentally different from that for a balanced and symmetrical one:
\begin{enumerate}
  \item In a general system, the degrees of the nodes should not be exactly proportional to their mean capacities.
  \item 
      For ``small nodes'', the ratios between their degrees and the total degree should be higher than those between the mean and total capacities.
\end{enumerate}
Based on these two insights, we provide an optimal flexibility design construction that still benefits from the simplicity of the WPC.

Our main contribution is two-fold. First, in terms of flexibility designs, we introduce a new \emph{thresholding} scheme. We treat nodes with capacities below the threshold as if they were just on the threshold, and then apply the WPC. We call this method the \emph{thresholded probabilistic construction (TPC)}. More specifically, we define the \emph{importance factor} for a supply node $u$, $q(u) \propto \max\left\{\Es(u), c/m \right\}$, and the \emph{importance factor} for a demand node $v$, $p(v) \propto \max\left\{\Ed(v), c/n \right\}$, for some appropriately chosen constant $c$. Then, a pair of nodes $(u,v)$ is linked with a probability $r(u,v) \propto q(u)p(v)$. It is clear that for a symmetrical system with  $\Es(u) =\frac{1}{m}$ and $\Ed(v)=\frac{1}{n}$, the TPC reduces to the WPC.
Most of our technical effort  is to show the optimality of the TPC for general production systems, that is, it requires only the average degree of $O(\ln(1/\eps))$ to achieve  $(1-\eps)$-optimality w.h.p.
As argued in \pref{sec:PPC} below, the ``thresholding'' step in our TPC solution is essential for our optimal construction.
From a practical viewpoint, when the mean supplies for different plants (and mean demands for different products) are close to each other, the performance of the TPC and the WPC are similar. The more heterogeneous are the mean supplies and  demands, the better the empirical performance of the TPC as compared with the WPC. Please refer to the simulation studies in Section \ref{sec:exp} in the main text and Section \ref{sec:add_exp} in the electronic companion (e-companion)  for more details.


A second technical contribution is that our analysis provides several new techniques for establishing generalized graph expansion properties. These may be useful for solving other problems that require expansion properties of  graphs with non-uniform degree sequences. In particular,  we first reduce the proof of the $(1-\epsilon)$-optimality w.h.p. to a few generalized expansion properties of graphs constructed using the TPC (see \pref{sec:graph_exp}). These generalized expansion properties extend the notion of ``probabilistic expanders" in \cite{Chen:15}. In fact, they can be viewed as a continuous generalization of the probabilistic expansion property, which is based on the cardinality of a set of nodes. In the symmetrical and balanced setting, to establish expansion properties from the WPC, the proof in \cite{Chen:15}  basically proceeds by two steps. First, by applying existing concentration inequalities, it shows that the probability of an arbitrary set of nodes  not expanding (i.e., not having many neighbors) is as small as an inverse exponential function of the number of nodes. Second, by applying a union bound over all (exponentially many) of  the sets, it shows that the probability that a non-expanding set exists remains very small. Therefore, the random construction via the WPC has the desired expansion property w.h.p. However, such a proof cannot work for asymmetrical and balanced systems  due to the following technical difficulties.  

When supply and demand are heterogeneous,  direct applications of existing concentration inequalities (e.g., Bernstein or Chernoff inequalities) provides loose upper bounds on the probability that an arbitrary set does not expand.  To resolve this problem, we prove a new concentration result via the exponential moment method and exploring the property of the moment generating function (see \pref{lem:Vs}). 

Moreover,  the probability that the neighboring set has a low total capacity (i.e., the set fails to expand) may not be as small as an inverse exponential function of the number of nodes. Thus, a direct union bound will fail. This happens because a set of nodes may
connect to a small number of nodes with high expected capacities rather than a great number
of nodes with small expected capacities. While the expected total capacities are roughly the same, a connection to fewer nodes does not guarantee the desired concentration rate. Let us look closely at this problem. When the realization of the supply or demand of a large node $u$ does not reach its expected capacity, many sets connected to $u$ do not have the desired generalized expansion property. 
When the union bound is applied  na\"ively, the probability that $u$ fails to reach its expected capacity (i.e., the intersection of the events that sets connected to $u$ fail to expand) is counted exponentially many times, and we cannot afford such double counting. To address this challenge, we carefully rearrange the events in a hierarchical way so that they do not overlap too much, which leads to a much tighter bound when applying the union bound  (see Section \ref{sec:general-case-overview} for details).

\subsection{Organization and  notations}
\label{sec:org}
The rest of this paper is organized as follows.  In \pref{sec:assump}, we introduce our assumptions about the general production systems considered  and additional necessary background.  In \pref{sec:construction}, we provide an optimal design. In particular, we first  show that the simple and intuitive WPC is sub-optimal in \pref{sec:PPC}. Then, in \pref{sec:TPC}, we introduce the thresholding idea to the WPC and propose our optimal construction: the TPC. Our analysis of the performance of the TPC is rather technical, so we first outline the proof in \pref{sec:framework}. In particular, we provide some generalized graph expansion properties that serve as  sufficient conditions of the desired $(1-\eps)$-optimality of the design. We  provide  further proofs of these generalized graph expansion properties in \pref{sec:proof}. The proofs of other results and technical lemmas, in addition to some concentration inequalities are provided in the e-companion.

Throughout the paper, we heavily use the asymptotic notations $O(\cdot)$, $\Omega(\cdot)$, $\Theta(\cdot)$, $o(\cdot)$, and $\omega(\cdot)$. Roughly speaking, $f(n)=O(g(n))$ means that $f$ is bounded above by $g$ (up to constant factor) asymptotically; $f(n)=\Omega (g(n))$ means that $f$ is bounded below by $g$ asymptotically; $f(n)=\Theta (g(n))$ means that $f(n)=O(g(n))$ and $f(n)=\Omega (g(n))$; $f(n)=o(g(n))$ means that $f$ is dominated by $g$  asymptotically; and $f(n)=\omega (g(n))$ means that $f$ dominates $g$ asymptotically.  Please also see  Chapter 3.1 in \cite{Cormen:2009} or \url{https://en.wikipedia.org/wiki/Big_O_notation} for rigorous definitions.  In addition, we  use the standard notation ``$\poly\log n$'' to denote some polynomial in $\log(n)$; similarly, ``$\poly n$'' means some polynomial in $n$.
{ In this paper, one should interpret the asymptotic notation on multiple variables $f(\eps, n, m) = O(g(\eps, n, m))$ as follows: there exists a universal constant $C$ so that $f$ is bounded above by $C\cdot g$ for sufficiently small $\epsilon$ and for every $n$ and $m$ satisfying the assumptions (see Assumption \ref{assumption:main-1} for details). Moreover, when we write $\zeta = \barn^{-\Omega(1)}$, it means  there exists some absolute constant $C > 0$ such that $\zeta \leq \barn^{-C}$ for sufficiently small $\epsilon$ and all $n$ and $m$ as required by the assumptions.
}

Throughout the paper, we use $c_0$, $c_1$, $c_2$, etc to denote absolute constants that do not depend on any other parameters. We further provide the computed values for all of these constants. 
To ease the understanding of the idea behind the proof, we suggest readers simply ignore the values of these constants when reading the proof.

\section{Assumptions and Background}
\label{sec:assump}

In this section, we first describe our assumptions about the general production systems.  Let us recall that  $d: V \rightarrow \R_{\geq 0}$  and $s: U \rightarrow \R_{\geq 0}$ denote the (possibly) random demand and supply function, respectively. The total expected supply and demand are normalized to one, i.e.,   $\sum_{u \in U}\Es(u) = \sum_{v \in V} \Ed(v) =1$.  Throughout this paper, we consider a general production system in which random supply and demand functions satisfy the following conditions (note that $\bar{n}=\max(m, n)$).
\begin{assumption} \label{assumption:main-1}

\begin{enumerate}
\item\label{enum:assumption-boundedness} $\kappa$-bounded variation with $\kappa \geq 1$: $\forall \; u\in U, v\in V$, $0\le s(u)\le \kappa \Es(u)$ and $0\le d(v)\le \kappa \Ed(v)$.
	\item  \label{enum:uniform-weak-ub} Upper bounds on expected supply and demand: { $\forall u\in U, v\in V, \Es(u)\le \frac{\cmx \eps^2}{\kappa^3  \ln \barn}, \Ed(v)\le \frac{\cmx \eps^2}{\kappa^3  \ln \barn}$ } for  some constant $\cmx<1$. 
    \item \label{enum:assumption-na} The random supply variables $s(u)$ for $u \in U$ are negatively associated and random demand variables $d(v)$ for $v \in V$ are  negatively associated.
\end{enumerate}
\end{assumption}

We now comment on our assumptions, which allows both random supply and non-\emph{i.i.d.} supply and demand. The first condition, which assumes that both $s(u)$ and $d(v)$ have a bounded variation of $\kappa$, is a standard assumption in the literature (see, e.g., \cite{Chou:11}, \cite{Chen:15}).

The second assumption on the expected supply and demand is rather weak, which allows \emph{highly heterogeneous} supplies and demands.  Take the demand side as an example. As $\sum_{v \in V} \Ed(v) = 1$ in a symmetrical system, we would have $\Ed(v) = \frac{1}{n}$ for each $v \in V$. In contrast, our condition on $\Ed(v)$ (i.e., $\Ed(v) = O\left(\frac{\epsilon^2}{ \ln \barn}\right)$) is exponentially looser than the symmetric case. { Moreover, our upper bounds on $\Es(u)$ and $\Ed(v)$ are necessary, as otherwise even the full flexibility system could not achieve $(1-\epsilon)$-optimality w.h.p. To see this, suppose that $n = m$ and the demands $d(v)$ are \emph{i.i.d.} over the first $o\left(\frac{\ln n}{\eps^2}\right)$ demand nodes and $d(v)=0$ for the rest nodes. This implies that { $\Ed(v) = \omega\left(\frac{\epsilon^2}{ \ln n}\right)$} for the first $o\left(\frac{\ln n}{\eps^2}\right)$ demand nodes. Based on the standard anti-concentration results, it is easy to see that, $\sum_{v \in V} d(v)$ will not concentrate to one w.h.p.  (although its expectation $ \sum_{v \in V} \Ed(v) =1$). In this case, even a fully flexible system can not guarantee that $(1-\epsilon)$ fraction of the total demand will be satisfied w.h.p. On the other hand, from a practical perspective, if  any node has an excessively large expected mean, we can simply add all of the edges to this node, which does not significantly increase the average degree.}
\label{page:assump}

The third assumption relaxes the independence condition. The negative association is a common assumption for modeling the correlation structure of multivariate distributions,  which subsumes the independence as a special case and  has a  wide range of applications (see, e.g., \cite{Shanthikumar:07}, \cite{Dubhashi:2009}). Let us recall the definition of negatively associated random variables. For $n$ random variables $X_1, \ldots, X_n$, they are said to be negatively associated if for every pair of disjoint subsets $A_1$ and $A_2$ of $\{1, \ldots, n\}$, $\text{Cov}[f(X_i, i \in A_1), g(X_j, j\in A_2)] \leq 0$ for all non-decreasing functions $f$ and $g$. Negatively associated multivariate distributions have many interesting properties \citep{JP83}. For example, the conditional distribution of $n$ independent random variables $X_1, \ldots, X_n$ given its sum $\sum_{i=1}^n X_i$ is negatively associated (Theorem 2.8 in \cite{JP83}). This property makes negative association a kind of realistic assumption for modeling random supplies and demands (e.g., considering the case that each demand node receives its own demand independently but the total demand is predetermined). { We also note that if the supply and demand variables are fully independent, then we do not need specialized concentration inequalities for negatively associated random variables (see Section \ref{sec:EC_tool} in the e-companion). Standard concentration inequalities for independent variables are sufficient for our purpose.}
\label{page:neg_associated}


We deal with not only heterogeneous supplies and demands,  but also unbalanced systems where the number of supply nodes $m$  can be significantly different from the number of demand nodes $n$. In fact, item \ref{enum:uniform-weak-ub} of \pref{assumption:main-1} implies the following relation between $m$ and $n$:
\begin{assumption} [Implied by \pref{assumption:main-1}] \label{assumption:main-2}
{ $\displaystyle{\frac{\min \{n, m\}}{\ln \barn}\ge \frac{\kappa^3}{ \cmx\epsilon^2 }}$.}
\end{assumption}
Here, $\kappa$ and $\cmx$ are the constants in \pref{assumption:main-1}. To see this implication,  based on the normalization of the expected total  supply and demand, there is a supply node $u \in U$  with $\bar{s}(u) \geq \frac{1}{m}$ and a demand node $v \in V$ with $\bar{d}(v) \geq \frac{1}{n}$. Based on item \ref{enum:uniform-weak-ub} of \pref{assumption:main-1}, we have { $\frac{1}{m} \leq \bar{s}(u) \leq \frac{\cmx \epsilon^2}{\kappa^3 \ln \barn}$ and $\frac{1}{n} \leq \bar{d}(v) \leq \frac{\cmx \epsilon^2}{\kappa^3 \ln \barn}$,} which further implies \pref{assumption:main-2}.  We also note that this assumption is not restrictive, as the gap between $n$ and $m$ can still be exponentially large (e.g., $n= e^{c' m}$ for some { $c' < \frac{\cmx \epsilon^2}{\kappa^3} $}).

In addition to the assumptions about production systems, we introduce some necessary background and notations for our theoretical development.
For any subset $L \subseteq U$ and $K \subseteq V$, we define
\[
s(L)\triangleq \sum_{u\in L}s(u), \quad d(K)\triangleq \sum_{v\in K}d(v),
\]
and
\[
\Es(L) \triangleq \E s(L) = \sum_{u\in L}\E s(u) = \sum_{u\in L}\Es(u),  \quad \Ed(K) \triangleq \E d(K) = \sum_{v\in K}\E d(v) = \sum_{v\in K}\Ed(v).
\]
From the normalization of the expected total  supply and demand, we have $\Es(U)=\Ed(V)=1$. Furthermore, for each $L\subseteq U$, we use $L^c \triangleq U\backslash L$ to denote the complement of $L$ with respect to $U$. With a slight abuse of notation, for each $K\subseteq V$,  $K^c \triangleq V \backslash K $. We have $\Es(L)+\Es(L^c)=1$ and   $\Ed(K)+\Ed(K^c)=1$ for any $L\subseteq U$ and $K\subseteq V$.

Given an undirected graph $G$ and a subset $K$ of the vertices of $G$, let $\Gamma_G(K)$ denote the neighborhood of $K$. When the underlying graph $G$ is clear from the context, we omit the subscript $G$ in $\Gamma_G(K)$. Based on the classical max-flow min-cut theorem,  the fulfilled demand with the realized supplies $\{s(u)\}_{u \in U}$ and demands $\{d(v)\}_{v \in V}$ can be written as,
\begin{equation}\label{eq:min_cut}
 \calZ_G(s,d)=\min_{L \subseteq U} \left\{\sum_{u \in L^c}s(u)+ \sum_{v \in \Gamma(L)}d(v)\right\}=\min_{L \subseteq U} \left\{s(L^c)+d(\Gamma(L))\right\}.
\end{equation}
Based on \eqref{eq:min_cut}, the goal of $(1-\eps)$-optimality w.h.p. in \eqref{eq:goal} can be equivalently stated as follow: with a probability of at least $1-\zeta$ (with $\zeta=\bar{n}^{-\Omega(1)}$), $s(L^c)+d(\Gamma(L)) \geq 1-\eps$ for any $L \subseteq U$.

We further show a simple fact:  under \pref{assumption:main-1}, the realized total supply $s(U)$ and demand $d(V)$ concentrate to $1$.
\begin{lemma}\label{lem:global_concentration} Under \pref{assumption:main-1}, with a high probability over $s(\cdot)$ and $d(\cdot)$,
	\begin{equation}
	1-\epsilon \leq s(U)\leq 1+\epsilon, \quad \textrm{ and } \quad 1-\epsilon \leq d(V)\leq 1+\epsilon.\label{eqn:global_concentration}
	\end{equation}	
\end{lemma}
The proof of \pref{lem:global_concentration} is a direct consequence of the Bernstein's inequality. (See its proof in Section \ref{sec:proof_full} in the e-companion.) From now on, for simplicity, we assume that \eqref{eqn:global_concentration} holds. More specifically, we condition our result on the event that \eqref{eqn:global_concentration} holds, which happens w.h.p. based on \pref{lem:global_concentration}.

Lemma \ref{lem:global_concentration} also suggests that the goal of $(1-\epsilon)$-optimality is achievable at least by the full flexibility design under  \pref{assumption:main-1}. To see this, note that the maximum flow of a complete bipartite graph is the minimum of the total supply and l demand, that is, $\calZ_F(s, d) = \min\{s(U), d(V)\}$ (where we let $F$ denote the design with full flexibility.). Then, based on \pref{lem:global_concentration} and by applying the union bound, we have w.h.p.:
\begin{equation}\label{eq:Z_F}
1 -\epsilon \leq \calZ_F(s, d) = \min\{s(U), d(V)\} \leq 1 + \epsilon.
\end{equation}

We note that our optimality criterion in \eqref{eq:goal} is slightly different from the common $(1-\epsilon)$-optimality criterion in the literature: $\calZ_G(s,d) \geq (1-\eps) \calZ_F(s,d)$. However, based on \eqref{eq:Z_F}, we have $1-\epsilon \leq \calZ_F(s,d) \leq 1+\epsilon$ w.h.p., and thus two optimality criteria are essentially equivalent. We choose the optimality criterion in \eqref{eq:goal} mainly for the ease of presentation.

\section{Construction}
\label{sec:construction}

The high-level framework of our thresholding probabilistic construction (TPC) is presented as follows. We associate a non-negative value $q(u)$ with each supply node $u \in U$ and a non-negative value $p(v)$ with each demand node $v \in V$, which represents their importance. The importance of the pair $(u,v)$ is defined as $q(u)p(v)$. We connect $(u,v)$ with a probability proportional to the importance, that is, with a probability $r(u, v) = \min\{ \gamma \barn q(u)p(v), 1\}$. The normalization factor $\gamma \barn $ is chosen to ensure that the resulting random graph achieves $(1 - \epsilon)$ optimality w.h.p. Under this framework, the key challenge is determining how to choose proper importance functions $q(u)$ and $p(v)$. We first provide a concrete example to show that a natural choice of importance functions in the WPC, that is, $q(u)=\bar{s}(u)$ and $p(v)=\bar{d}(v)$, would fail.


\subsection{Sub-optimality of the weighted probabilistic construction}
\label{sec:PPC}

In this subsection, we study the weighted probabilistic construction (WPC), where $q(u) = \Es(u)$ and $p(v) = \Ed(v)$. 
We prove the following theorem, which shows that the WPC is a sub-optimal flexibility design. 

\begin{theorem}\label{thm:WPC-suboptimal}
For any $\epsilon \in (0, 1)$ and any $\alpha \in \left(4\epsilon, \frac{1}{\ln(1/\epsilon)}\right)$, there is a family of balanced systems ($n = m$ and $n$ is an even number for simplicity) such that for each system in the family, the WPC method needs $\Omega(n/\alpha)$ edges to achieve $(1-\epsilon)$-optimality w.h.p.
\end{theorem}

\proof{Proof.}
We first construct the system for every even integer $n$. The supplies are deterministic, i.e., $s(u) = \Es(u)$ is a constant for each $u \in U$. The supply nodes are not uniform, and can be split into two equal-sized subsets $U_1$ and $U_2$, each with $\frac{n}{2}$ nodes. We set $\Es(u) = \frac{2 - \alpha}{n}$ for each $u \in U_1$ and set $\Es(u) = \frac{\alpha}{n}$ for each $u \in U_2$. The demands are \emph{i.i.d.} following a two-point distribution:
\[
d(v) = \left\{
\begin{array}{cc}
0 & \mbox{with probability}~1/2\\
2\Ed(v) & \mbox{with probability}~1/2
\end{array}
\right. , \footnote{Note that one can make small modifications to this construction so that the factor $2$ before $\Ed(v)$ becomes an arbitrary constant that is strictly greater than $1$.}
\]
with $\Ed(v) = \frac{1}{n}$ for every $v \in V$. It is clear Assumption \ref{assumption:main-1} holds for this instance.

The probability of connecting $u$ and $v$ by an edge in the WPC is
\[
\gamma n \Es(u) \Ed(v)= \left\{
\begin{array}{cc}
\gamma(2-\alpha) n^{-1} & \mbox{if}~u \in U_1\\
\gamma \alpha n^{-1} & \mbox{if}~u \in U_2\\
\end{array}
\right. ,
\]
and the expected number of edges of the construction is
\[
\sum_{u \in U, v \in V} \gamma n \Es(u) \Ed(v) =  \gamma n \Es(U) \Ed(V)= \gamma n .
\]
We now argue that we need $\gamma = \Omega(\frac{1}{\alpha})$  to achieve $(1-\epsilon)$-optimality, making the total number of edges $\Omega(n/\alpha)$. To see this, let us suppose the contrary. If $\gamma < \frac{1}{4\alpha}$, the expected number of edges incident to $U_2$ is $\gamma \alpha n^{-1} \cdot n^2/2 < n/8$. Therefore, by Markov inequality, with a probability of at least $\frac12$ there will be fewer than $n/4$ edges incident to $U_2$, leaving more than $n/4$ nodes in $U_2$ disconnected from every demand node. All the disconnected supply nodes in $U_2$ cannot be consumed. Therefore, more than $(n/4) \cdot \Es(u)_{|u \in U_2} =(n/4) \alpha n^{-1}> \epsilon$ supply cannot be consumed, and thus $(1-\epsilon)$-optimality cannot be achieved. \qed
\endproof

However, if we connect every $u \in U_1 \cup U_2$ to $100 \ln (1/\epsilon)$ random nodes in $V$, a straightforward adaptation of the analysis in \cite{Chen:15} shows that the constructed flexibility design achieves $(1-\epsilon)$-optimality w.h.p.  This design uses only $O(n\ln(1/\epsilon))$ edges, rendering the WPC with the importance functions $q(u) = \Es(u)$ and $p(v) = \Ed(v)$ sub-optimal.

\begin{remark}\label{rem:counter}
As long as $\alpha = o\left(\frac{1}{\ln(1/\epsilon)}\right)$, \pref{thm:WPC-suboptimal} states that the  WPC needs $\omega(n \ln (1/\epsilon))$ edges to achieve $(1-\epsilon)$-optimality w.h.p. For example, when $\alpha=\sqrt{\epsilon}$, the WPC requires $\Omega(n/\sqrt{\epsilon})$ edges to achieve $(1-\epsilon)$-optimality w.h.p.  It is worthwhile to note that  the condition on $\alpha$ for the failure of the WPC does not require some supplies to be extremely small, which implies that the failure scenario is not unrealistic. For example, when $\epsilon=0.01$ (i.e.,  the goal is to achieve 99\% of the maximum flow of the full flexibility) and $\alpha=\sqrt{\epsilon}=0.1$, the two levels of (normalized) supplies are $1.9/n$ and $0.1/n$, respectively, according to the proof of Theorem \ref{thm:WPC-suboptimal}. In this failure scenario,  small supplies are not negligible as compared with large supplies. Our experimental result shows that the TPC improves significantly over the WPC in this case (see Figure \ref{fig:two_1}).
\end{remark}

Furthermore, although the nodes in $U_2$ have much less capacity than the nodes in $U_1$, their degrees should be as high as $\Omega(\ln(1/\epsilon))$. For exemplary purposes, let us fix $\alpha = \sqrt{\epsilon}$, and show that at least half of the nodes in $U_2$ should have a degree greater than $\ln(1/\epsilon)/8$. Suppose for contradiction that more than half of the nodes in $U_2$ have at most $\ln(1/\epsilon)/8$ neighbors in $V$. For each of such nodes in $U_2$, with a probability $2^{-\ln(1/\epsilon)/8} > \epsilon^{1/4}$, none of its neighbor has positive demand. Therefore, in expectation, there are at least  $\epsilon^{1/4} \cdot \frac{|U_2|}{2} = \frac{\epsilon^{1/4} n}{4}$ nodes in $U_2$ with no positive-demand neighbor, and their supply cannot be consumed. Therefore, we   lose $\frac{\epsilon^{1/4} n}{4} \cdot \frac{\alpha}{n} = \frac{\epsilon^{3/4}}{4}$ supply in expectation, and thus cannot achieve $(1-\epsilon)$-optimality w.h.p. (for small  $\epsilon$).

\subsection{Thresholded probabilistic construction}
\label{sec:TPC}

In this section, we present the proposed optimal construction, the TPC, based on a novel choice of the importance functions.

The example discussed in \pref{sec:PPC} suggests that the importance of a node should be significantly higher than its mean capacity when its mean capacity is very small. Inspired by this implication, we raise the importance of a node if its mean capacity is less than a threshold of $O(\frac{1}{m})$ (for a supply node) or $O(\frac{1}{n}$) (for a demand node). Formally, for each supply node $u \in U$ and each demand node $v \in V$, we define the importance functions
\begin{equation}\label{eq:pq-construction}
q(u) = \frac{1}{n_q}\max\left\{\Es(u), \frac{1}{\cp  m}\right\} ~~\mbox{and}~~ p(v) = \frac{1}{n_p}\max\left\{\Ed(v), \frac{1}{\cp  n}\right\},
\end{equation}
 where $n_p$ and $n_q$ are normalization factors, so we have
$\sum_{u \in U} q(u) = 1$ and $\sum_{v \in V} p(v) = 1,$
that is, $n_q =\sum_{u\in U}\max\left\{\Es(u), \frac{1}{\cp  m}\right\}$ and $n_p =\sum_{v\in V}\max\left\{\Ed(v), \frac{1}{\cp  n}\right\}.$


For notational convenience, we also extend the definition of $p(\cdot)$ and $q(\cdot)$ to the domain of all subsets:  
\[q(L)\triangleq \sum_{u\in L}q(u), \;\; \forall \; L\subseteq U ~~\mbox{and}~~ p(K): = \sum_{v\in K}p(v), \;\; \forall \; K\subseteq V.
\]
It is worth noting that $p(\cdot)$ and $q(\cdot)$ are deterministic functions on subsets of $U$ and $V$, respectively, which are lower-bounded by $\Es(\cdot)$ and $\Ed(\cdot)$ up to a constant factor, respectively. Moreover, $p(V)$ and $q(U)$ are normalized at 1. We summarize the properties of $p$ and $q$ in the following proposition:
\begin{proposition}\label{prop:dprelation}
Let us define the constant $\ck= \ckdisp$. We have $n_p \le \frac{1}{\ck}$ and $n_q \le \frac{1}{\ck}$, so
\begin{equation*}
p(v)\ge \ck \cdot\Ed(v) \quad \emph{and} \quad q(u)\ge \ck \cdot \Es(v)\quad \forall u\in U, v\in V.
\end{equation*}
Moreover, $p(V) = q(U)  =1$.
\end{proposition}
\proof{Proof.}  Based on the definition of $n_p$, we have
$
    n_p \leq \sum_{v \in V} \Ed(v)+ n \cdot \frac{1}{5\kappa n } = 1 + \frac{1}{5 \kappa} \leq \frac{6}{5}.
$
  The upper bound for $n_q$ can be established in a similar way. That $p(V) = q(U) = 1$ follows straightforwardly from the normalization.   \qed
  \endproof

We formally describe our design as follows. We use the following random process to generate a bipartite graph $G(U \cup V, E)$, which serves as the process flexibility design. We further denote the corresponding distribution of $G$ by $\mathcal{G}$. 

\bigskip
\noindent\fbox{%
\centering
    \parbox{0.98\textwidth}{%
\noindent{\bf Design from TPC: } For any pair of nodes $(u,v)\in U\times V$, we include $(u,v)$ into the edge set $E$ of $G$ with the probability
\begin{eqnarray}\label{eqn:tpc}
r(u,v) = \min\{\gamma \barn q(u)p(v),1\}
\end{eqnarray}
with $\gamma= \cgamma\kappa^3\ln(e\kappa)\ln(\frac{4\kappa}{\epsilon})$, where $c_0$ is an absolute constant.
}
}
\smallskip


Theoretically, the constant $\cgamma$ (in $\gamma$) can be set to $\cgammav$ according to our proof. However, constants independent of $\bar{n}$ and $\epsilon$ are not our main focus. The number of edges used in the TPC is small in expectation and concentrates to its expectation. In fact, as shown in the next proposition, the constructed design has an average degree of $O(\ln (1/\epsilon))$.
\begin{proposition}\label{prop:number-of-edges}
	For a design $G = (U \cup V, E) \sim \mathcal{G}$ from the TPC,
    \begin{align*}
    	\E |E| = O_{\kappa}\left(\barn \ln\left(1/{\epsilon}\right)\right),
    \end{align*}
    where { $O_{\kappa}(\cdot)$ hides a factor that depends only on the constant $\kappa$}. Furthermore, we have $|E| \leq 2 \E |E|$ w.h.p.
\end{proposition}
\proof{Proof of \pref{prop:number-of-edges}.}
  According to the TPC and the choice of $\gamma$, the expected number of edges is bounded from the preceding by
  \begin{equation}\label{eq:number_edge}
    \E |E|  =\sum_{u \in U} \sum_{v \in V} r(u,v) \le \gamma \barn = O_{\kappa}\left(\barn \ln \left(1/{\epsilon}\right)\right).
  \end{equation}
  Furthermore, based on the standard Chernoff bound,
$
 \Pr\left[|E| < 2\E|E| \right] \geq 1 - \exp\left(-\frac{\E|E|}{3}\right) = 1 - \barn^{-\omega(1)}.$
 \qed
\endproof

{
\begin{remark}\label{rem:UPC}
Our TPC can be viewed as a combination of the WPC and uniform probabilistic construction (UPC) as introduced in \cite{Chen:15} for balanced and symmetrical systems. More precisely, for properly chosen edge densities, we apply both the WPC and UPC. The union of two constructions receives a guarantee similar to that of our TPC.
\end{remark}
}

\subsection{Simulation study for the effectiveness of the TPC}
\label{sec:exp}

Before we theoretically prove that the TPC is an optimal construction in the next section, let us provide some simulation studies to illustrate its effectiveness as compared with the classical WPC. Let us first consider the setting in the proof of Theorem \ref{thm:WPC-suboptimal}, where the (normalized) deterministic supplies take the value $\frac{2-\alpha}{n}$ for the first half of the nodes and the value $\frac{\alpha}{n}$ for the remaining half. The demands are \emph{i.i.d.} with a two-point distribution and a mean of $\frac{1}{n}$. We choose different values of $\alpha\in \{0.1, 0.2, 0.3, 0.4\}$. For each $\gamma$ (i.e., the average degree), we construct 100 random designs using the TPC and 100 random designs using the WPC. We also generate 1,000 demand realizations. For each design $G$ and realization of the demand $d$, we compute the \emph{ratio} between the maximum flow of $G$ and that of the full flexibility $F$: $\frac{\mathcal{Z}_G(s,d)}{\mathcal{Z}_F(s,d)}$. In the TPC, we use a slightly different threshold from \eqref{eq:pq-construction} for better empirical performance, that is, $q(u) \propto \max\left\{\Es(u), \frac{c}{m}\right\}$ and $p(v) \propto \max\left\{\Ed(v), \frac{c}{n}\right\}$ with $c=0.5$. We note that the constant in the threshold does not affect our theoretical analysis, and we use the constant $c:=\frac{1}{5}$ in \eqref{eq:pq-construction} only for ease of calculation in some concentration inequalities. { In practice, when using a large constant $c$, the number of edges also increases by a constant factor. However, a smaller $c$ carries the risk of missing some random supplies/demands with small mean capacities. Thus, the threshold parameter $c$ controls the robustness vs. the scarcity. In practical scenarios, if some prior knowledge of supply/demand distributions exists, then the threshold parameter $c$ can be tuned by offline simulations.}

\begin{figure}[!t]
\centering
\subfigure[$\alpha=0.1$]{
  \includegraphics[width=0.45\textwidth, height=5.7cm]{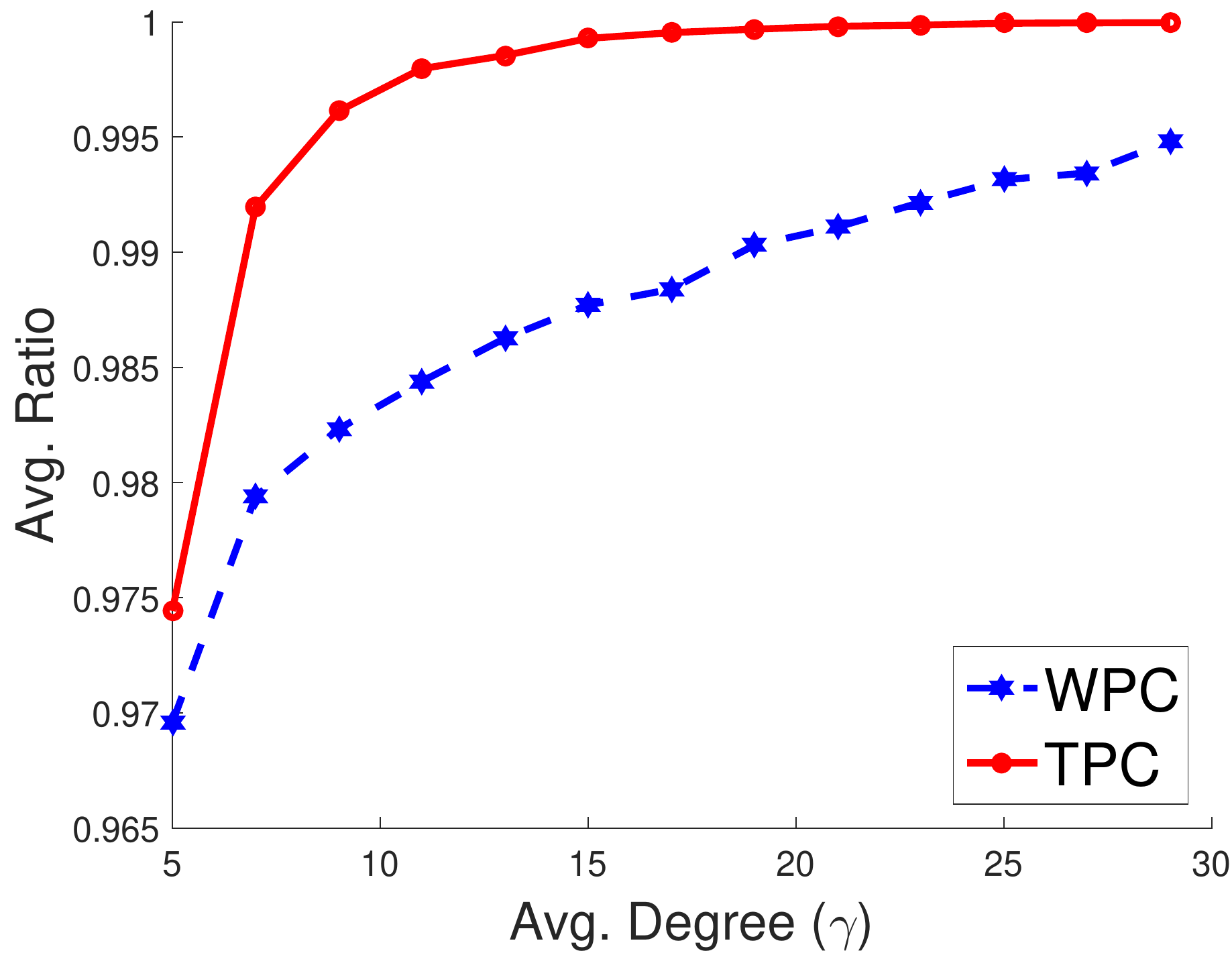}
	    \label{fig:two_1}
}
\subfigure[$\alpha=0.2$]{
  \includegraphics[width=0.45\textwidth,  height=5.7cm]{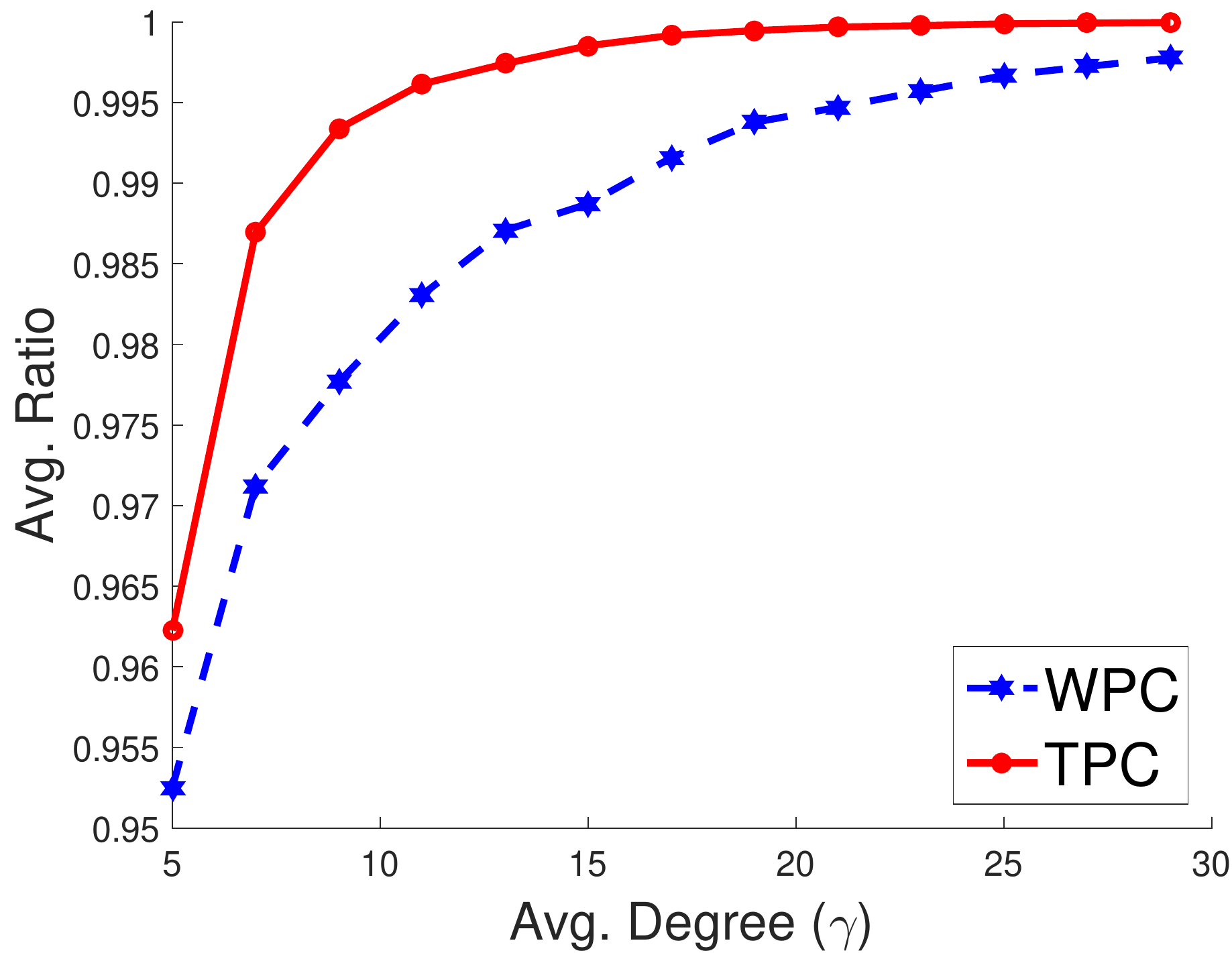}
	    \label{fig:two_2}
}
\subfigure[$\alpha=0.3$]{
  \includegraphics[width=0.45\textwidth,  height=5.7cm]{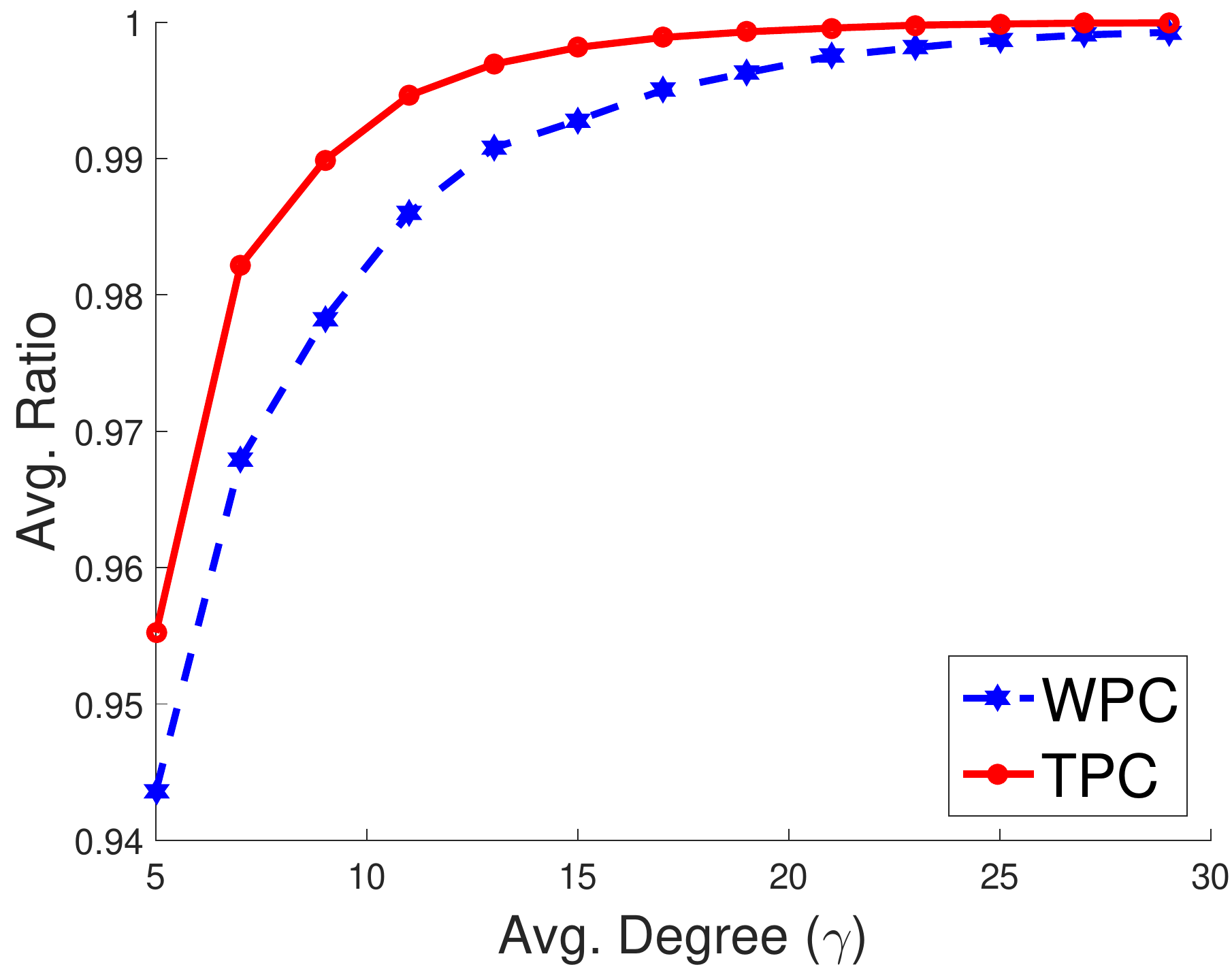}
	    \label{fig:two_3}
}
\subfigure[$\alpha=0.4$]{
  \includegraphics[width=0.45\textwidth,  height=5.7cm]{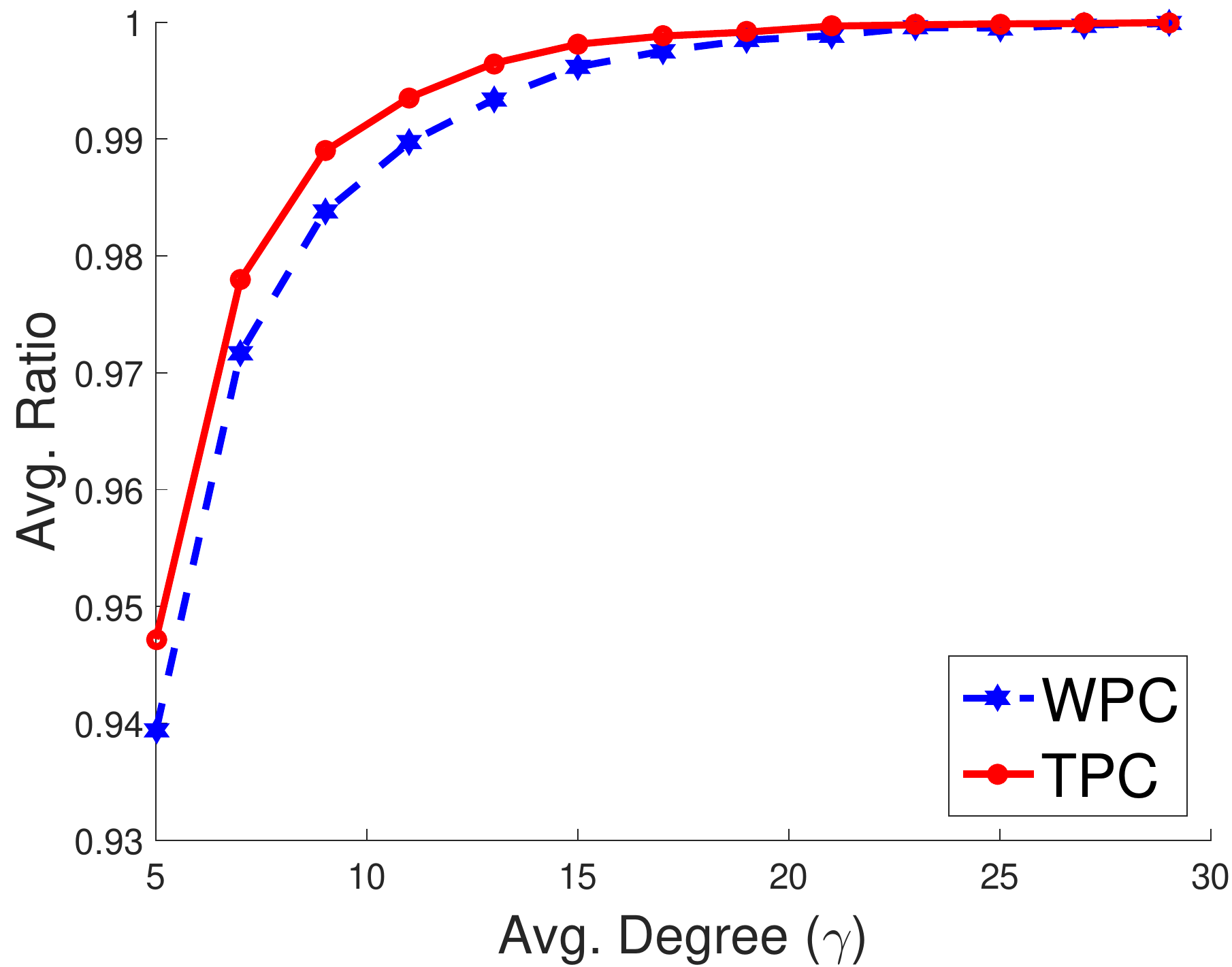}
	    \label{fig:two_3}
}
\caption{The comparison between the WPC and TPC. The $x$-axis is the average degree $\gamma$, which varies from 5 to 30. The $y$-axis is the averaged ratios between the maximum flow of the design $G$ and that of the full flexibility $F$ (the larger the better). Each ratio plotted in the graph is averaged over 100 random graphs for a given $\gamma$ and 1,000 demand realizations. }
\label{fig:comp_two}
\end{figure}

\begin{figure}[!t]
\centering
\subfigure[WPC design]{
  \includegraphics[width=0.45\textwidth, height=5.7cm]{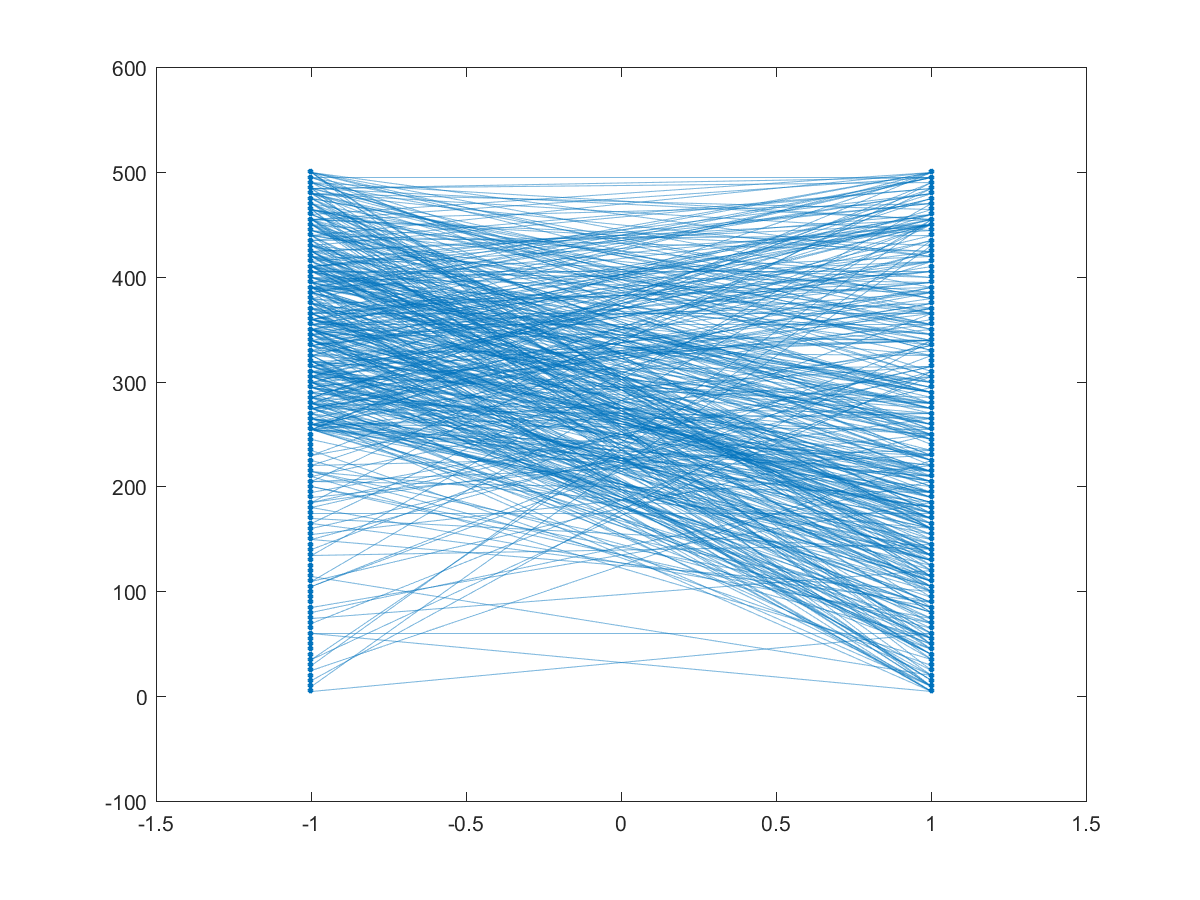}
	    \label{fig:WPC_two}
}
\subfigure[TPC design]{
  \includegraphics[width=0.45\textwidth,  height=5.7cm]{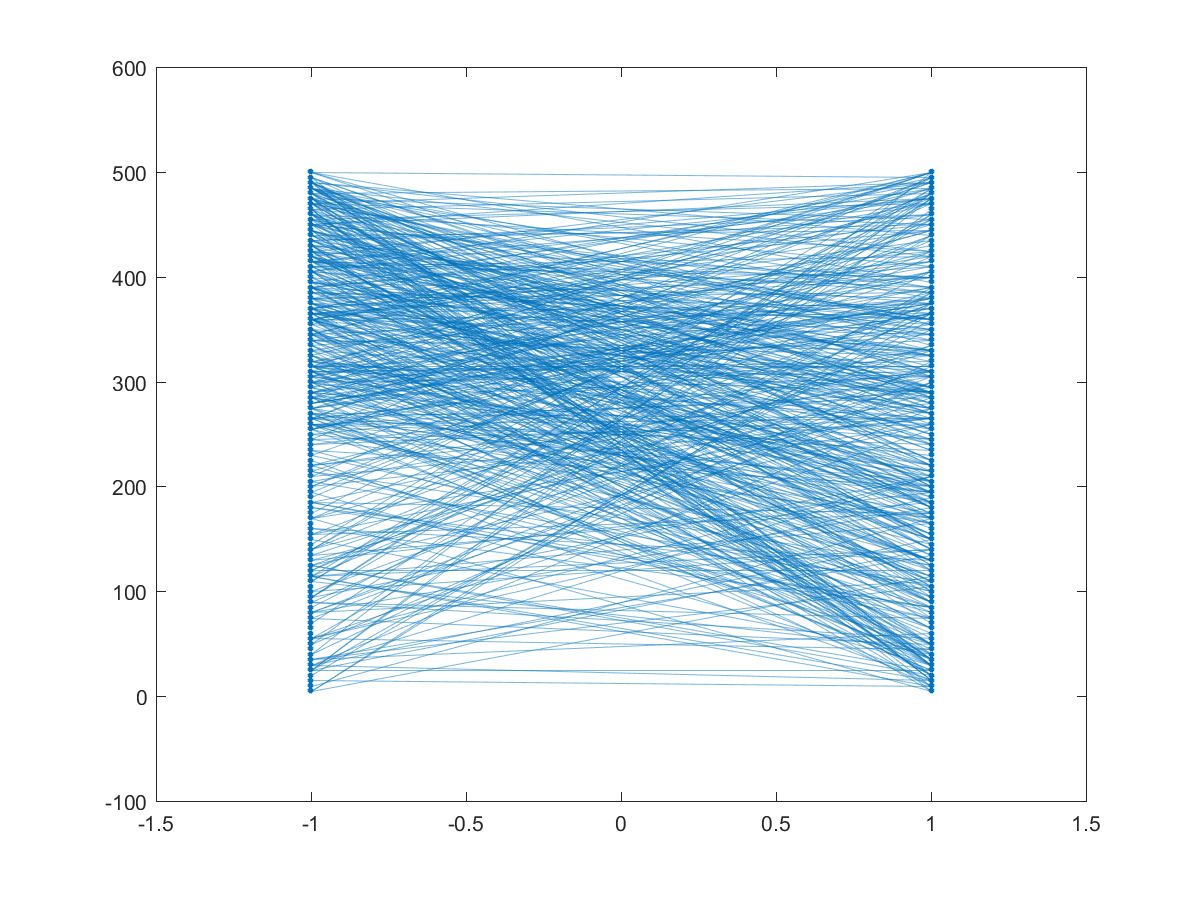}
	    \label{fig:TPC_two}
}
\caption{The designs using the WPC and TPC when $\gamma=5$ and $\alpha=0.2$}
\label{fig:two_graphs}
\end{figure}

We set $n=m=100$ in the experiment and present the averaged ratios in Figure \ref{fig:comp_two}. In Figure \ref{fig:comp_two}, the TPC clearly achieves improved performance over the WPC, especially when $\alpha$ is small (i.e., the supplies are more heterogeneous). With the averaged degree $\gamma$ of about 10, the TPC achieves more than 99\% of the maximum flow of the full flexibility. This simulation result matches the intuition in the proof of Theorem \ref{thm:WPC-suboptimal}, which shows that the nodes with small mean capacities need more edges. To see that, we plot two random designs using the WPC and TPC. Due to the thresholding scheme in the TPC, the number of edges connected to small supply nodes (i.e., the second half of the supply nodes) in the TPC is much larger than that in the WPC. In particular, in the TPC design in Figure \ref{fig:TPC_two}, about 20\% of the edges are connected to small supply nodes; while in the WPC design in Figure \ref{fig:WPC_two}, only about 10\% of the edges are connected to small supply nodes. 

For a wide range of heterogeneous supply and demand models, one can easily observe the improvement of the TPC over the WPC. In Section \ref{sec:add_exp} in the e-companion, we present other simulation studies when the mean capacities are drawn from a uniform distribution or power laws. In all of these settings, our experiments show that the TPC outperforms the WPC consistently and that the superiority becomes more noticeable when the mean capacities are more heterogeneous. For mean supplies/demands with a higher degree of heterogeneity, the thresholding scheme in the TPC is more effective, making more edges connected to small capacity nodes.

\section{Main Theoretical Result --- Optimality of the TPC}
\label{sec:framework}

In this section, we first introduce our main theorem (\pref{thm:main}) on the optimality of the TPC construction introduced in \pref{sec:TPC}, and reduce the proof of the optimality to a few generalized graph expansion properties of the obtained random design.

\begin{theorem}[Main]\label{thm:main}
	Assume that $\epsilon < 1/3$. With a high probability over the choice of $G = (U \cup V, E) \sim \calG$, we have
	\begin{enumerate}
		\item The number of edges in $E$ is $O_{\kappa}(\barn\ln(1/\eps))$, where $\barn = \max\{m,n\}$, { and $O_{\kappa}(\cdot)$ hides a factor that depends only on the constant $\kappa$};
		\item $G$ achieves $(1 - 2\eps)$ optimality w.h.p.
	\end{enumerate}
\end{theorem}

We note that the $(1 - 2\eps)$ optimality (instead of $(1-\eps)$ optimality) of $G$ merely facilitates presentation of the proof and one can always introduce $\eps'=2\eps$. According to the lower bound in \cite{Chen:15}, the proposed TPC leads to an optimal design (i.e., the most sparse design up to a constant factor) to achieve $(1-\epsilon)$ optimality w.h.p. In particular, Corollary 2 in \cite{Chen:15} shows that in a balanced and symmetrical system, the number of edges must be at least $\Omega(\barn\ln(1/\eps))$ to achieve $(1-\epsilon)$ optimality w.h.p. As we consider a more general system that subsumes a balanced and symmetrical system as a special case, the lower bound in \cite{Chen:15} automatically serves as a lower bound on the requirement of the number of edges to achieve $(1-\epsilon)$ optimality w.h.p.

The first statement of \pref{thm:main} follows directly from \pref{prop:number-of-edges} and { the} constant depending on $\kappa$ that hides in $O_{\kappa}(\cdot)$ is from \eqref{eq:number_edge}. The key is to prove the second statement on the optimality, that is, w.h.p. over the choice of $G \sim \mathcal{G}$, $G$ achieves $(1 - 2\eps)$ optimality w.h.p. over the randomness of $s(\cdot)$ and $d(\cdot)$. This claim can be mathematically stated as
\begin{equation}\label{eq:G_goal}
  \Pr_{G \sim \mathcal{G}}\left[ \Pr_{s(\cdot), d(\cdot)} \left( \mathcal{Z}_{G}(s,d) \geq 1-2\epsilon\right) \geq 1-\zeta \right] \geq 1-\zeta,
\end{equation}
for some $\zeta=\bar{n}^{-\Omega(1)}$.


{ Using a cut condition from the max-flow min-cut theorem (see \pref{lem:max-flow-min-cut-goal}), we are able to reduce the second goal of \pref{thm:main} to a pair of generalized expansion properties; see \pref{thm:starting_point}, \pref{lem:lefttorightexpansiondemand}, and \pref{lem:lefttorightexpansionp} in Section \ref{sec:graph_exp}. We then show that the obtained random graph $G$ using the TPC will satisfy these generalized expansion properties w.h.p. This high-level idea is similar to the approaches in \cite{Chou:10,Chou:11} and \cite{Chen:15}. However, the novel part of the proof is to show that the generalized expansion properties are satisfied w.h.p. As mentioned in the introduction, due to the heterogeneous supplies and demands, existing concentration inequalities will lead to loose upper bounds and a direct use of union bound will fail. We overcome these difficulties by developing new technical tools in Section \ref{sec:proof}.
}

\subsection{From $(1-\epsilon)$ optimality to generalized graph expansion  properties}
\label{sec:graph_exp}



In this section, we reduce the proof of \pref{thm:main} to a set of generalized expansion properties. To this end, we first state the following lemma, which is a direct application of the max-flow min-cut theorem in \eqref{eq:min_cut}.
\begin{lemma}\label{lem:max-flow-min-cut-goal}
Given a realization of the supply vector $s(\cdot)$ and demand vector $d(\cdot)$, the fulfilled demand (i.e., the maximum flow) is at least $(1 - 2\epsilon)$ if and only if for any $L\subseteq U$,
\begin{equation}
d(\Gamma(L))+ s(L^c) \ge 1-2\epsilon. \label{eqn:goal}
\end{equation}
\end{lemma}
Based on \pref{lem:max-flow-min-cut-goal}, the second statement of \pref{thm:main} reduces to prove that \eqref{eqn:goal} holds w.h.p. over the choice of $G$ and the realizations of $s(\cdot)$ and $d(\cdot)$.

Now let us place a condition on the event that the equation \eqref{eqn:global_concentration} holds (i.e., $1-\epsilon\leq s(U) \leq 1+\eps$ and $1-\eps \leq d(V) \leq 1+\eps$), which happens w.h.p. based on \pref{lem:global_concentration}. For a fixed $L \subseteq U$, to establish \eqref{eqn:goal}, it suffices to show that
 \begin{equation}
		d(\Gamma(L)) \ge s(L) -\epsilon, \label{eqn:slargerd}
\end{equation}
To see this, we note that \eqref{eqn:slargerd} implies $d(\Gamma(L))+ s(L^c)  \geq s(L)+ s(L^c)  -\epsilon = s(U) -\epsilon  \geq 1-2\epsilon$, which gives \eqref{eqn:goal}.

Meanwhile, let $K = V\setminus \Gamma(L)$. To establish \eqref{eqn:goal}, it also suffices to show that
\begin{equation}
			s(\Gamma(K))\ge d(K) -\epsilon,
 \label{eqn:dlargers}
\end{equation}
as
\begin{align*}
s(\Gamma(K))\ge d(K) -\epsilon \Rightarrow \; & s(\Gamma(V \setminus \Gamma(L)))\ge d(V) - d(\Gamma(L)) -\epsilon & & \mbox{(based on the definition of $K$)}\\
\Rightarrow \; &d(\Gamma(L)) + s(L^c) \geq d(V) - \epsilon && \mbox{(as $L^c \supseteq \Gamma(V \setminus \Gamma(L))$)}\\
\Rightarrow \; &d(\Gamma(L)) + s(L^c) \geq 1 - 2\epsilon. && \mbox{(based on \eqref{eqn:global_concentration})}
\end{align*}

In summary, to establish \eqref{eqn:goal}, we can choose to prove either of \eqref{eqn:slargerd} and \eqref{eqn:dlargers}, whichever is easier. In other words, to prove the second statement of \pref{thm:main}, it suffices to prove the following theorem.

\begin{theorem}\label{thm:starting_point}
With high probability over the choice of graph $G$ and supply and demand functions, for any subset $L\subseteq U$, either \eqref{eqn:slargerd} or \eqref{eqn:dlargers} holds.
\end{theorem}

{ The formal proof of the reduction from \pref{thm:main} to \pref{thm:starting_point} is provided in Section \ref{sec:reduction} in the e-companion.} 
We now explain that \eqref{eqn:slargerd} and \eqref{eqn:dlargers} are generalizations of classical graph expansion properties (see, e.g., \cite{Hoory:06} and the reference therein) and the probabilistic expansion property proposed in \cite{Chen:15}. For ease of illustration, we temporarily drop the normalization assumption about $s(\cdot)$ and $d(\cdot)$ in the following discussion. Recall that given a bipartite graph $G=(U \cup V, E)$, the expansion property from $U$ to $V$ says that for any not-too-large set $L \subseteq U$, the size of its neighbor will be at least $\lambda |L|$ for some constant $\lambda$, that is, $|\Gamma_G(L)| \geq \lambda |L|$. One can similarly define the expansion property from $V$ to $U$. In \eqref{eqn:slargerd} and \eqref{eqn:dlargers}, we obtain similar expansion criteria when $s(\cdot)$ and $d(\cdot)$ are constant functions. For example, if $s(\cdot)$ is set to a constant--$1$ function and $d(\cdot)$ is set to a constant--$\frac{1}{2}$ function, then \eqref{eqn:slargerd} is equivalent to $|\Gamma(L)| \geq 2 |L| - 2\epsilon$, which corresponds to $\lambda = 2$ in the expansion property if we ignore the $-2\epsilon$ term. When $d(\cdot)$ is set to the constant--$1$ function and each $s(u)$ is an independent Bernoulli random variable such that $s(u) = 0$ with a probability of $\frac{1}{2}$ and $s(u) = 2$ with a probability of $\frac{1}{2}$, we define the random set $T = \{u \in U: s(u) = 2\}$. We observe that \eqref{eqn:slargerd} is equivalent to $\forall L \subseteq U, ~ |\Gamma(L)| \geq 2|L \cap T| - \epsilon$. This condition is further equivalent to
\begin{equation}\label{eq:explain-expanding-property-relation}
\forall L' \subseteq T, |\Gamma(L')| \geq 2|L'| - \epsilon,
\end{equation}
noting that $|\Gamma(L)| \geq |\Gamma(L\cap T)|$ for all $L \subset U$.
The property that \eqref{eq:explain-expanding-property-relation} holds for a random set $T$ w.h.p. is essentially the probabilistic expansion property introduced in \cite{Chen:15}. A similar comparison can be made for \eqref{eqn:dlargers}. As $s(\cdot)$ and $d(\cdot)$ considered in this paper are general continuous random functions, the properties \eqref{eqn:slargerd} and \eqref{eqn:dlargers} can be viewed as generalizations of the probabilistic expansion property in \citet{Chen:15}.

{ To prove Theorem \ref{thm:starting_point}, we further reduce the expansion properties in \eqref{eqn:slargerd} and \eqref{eqn:dlargers} to the expansion properties involving the sampling probabilities $p(\cdot)$ and $q(\cdot)$ defined in \eqref{eq:pq-construction}. In particular, we prove Theorem \ref{thm:starting_point} by introducing the following two lemmas.}

\begin{lemma}\label{lem:lefttorightexpansiondemand}
	With high probability over the choice of $G$ and the supply and demand functions $s(\cdot)$ and $d(\cdot)$, for any $L \subseteq U$ with $\ck \epsilon/\kappa \le q(L)$, we have
	\begin{equation}
	d(\Gamma(L))\ge \min\left\{1-\delta, \; \frac{\kappa}{\ck} q(L)\right\}\label{eqn:inter13}.
	\end{equation}
	where 
	$\delta = \frac{1}{\deltav}$,
	$\ck = \ckdisp$ are two constants, and $\kappa$ is defined in \pref{assumption:main-1}.
\end{lemma}
Note that the constant $\ck$ in \pref{lem:lefttorightexpansiondemand} comes from \pref{prop:dprelation}.

\begin{lemma}\label{lem:lefttorightexpansionp}
	Assume that $\epsilon < 1/3$. With high probability over the choice of $G$, 
	for any $L \subseteq U$ with $q(L)\ge \tau$, we have
	\begin{equation}
	p(\Gamma(L)) \ge 1-\tau, \label{eqn:lefttorightexpansionnonlinear}
	\end{equation}
	where 
	$\tau = \frac{1}{2\kappa}$.
\end{lemma}

Both lemmas show expansion-like properties of a random graph $G$. In contrast to normal expansion properties using the set cardinality to measure a set, we use $p(\cdot)$, $q(\cdot)$, and $d(\cdot)$ to define the measure of a set $L$ and the associated $\Gamma(L)$. { The proof of Theorem \ref{thm:starting_point} using \pref{lem:lefttorightexpansiondemand} and \pref{lem:lefttorightexpansionp} is provided in Section \ref{sec:proof_of_starting} in the e-companion.}

\section{Proof of the Generalized Expansion Properties (\pref{lem:lefttorightexpansiondemand} and \pref{lem:lefttorightexpansionp})}
\label{sec:proof}

We now need to prove \pref{lem:lefttorightexpansiondemand} and \pref{lem:lefttorightexpansionp} to complete the proof of our main theorem (\pref{thm:main}). Let us first build up some basics for proving \pref{lem:lefttorightexpansiondemand}. \pref{lem:lefttorightexpansionp} will become much easier to prove once \pref{lem:lefttorightexpansiondemand} is established.

We first develop a useful functional form for $d(\Gamma(L))$ for a given $L \subseteq U$. Let $I_L(v)$ be the shorthand for the indicator variable for the event $v\in \Gamma(L)$, that is, $I_L(v) = 1$ when $v \in \Gamma(L)$, and $I_L(v) = 0$ otherwise. We omit the subscript $L$ in $I_L(v)$ when it is clear from the context.
Our key quantity $d(\Gamma(L))$ can then be written as the sum of $n$ negatively associated random variables: 
\begin{align} \label{eqn:dgammaL}
d(\Gamma(L)) = \sum_{v\in V} d(v)\mathbf{1}_{v\in \Gamma(L)} = \sum_{v\in V} d(v)I_L(v).
\end{align}
 Note that $\{d(v)\}_{v\in V}$ and $\{I_L(v)\}_{v \in V}$ are two independent sets of random variables. Based on \pref{prop:NA-union} in the e-companion, $\{ d(v)I_L(v)\}_{v \in V}$ are negatively associated.
We analyze our construction of the flexibility design to unveil the property of the indicator random variable $I_L(v)$ and to prove the concentration of $d(\Gamma(L))$ using \eqref{eqn:dgammaL}.
%
%

According to our TPC of the flexibility design (see \eqref{eqn:tpc}), we have
\begin{align}\label{eqn:inter11}
\Pr[I_L(v) = 1] & = 1- \Pr[v \not \in \Gamma(L)] = 1 - \prod_{u\in L}\Pr[(u,v)\not\in E] = 1 - \prod_{u\in L}(1-r(u,v)) \\
& {\geq}   1 - \prod_{u\in L}\exp\left(-\gamma q(u)p(v)\barn \right)
= 1 - \exp\left(-\gamma p(v)\cdot \left( \sum_{u\in L} q(u)\right) \cdot \barn \right) \nonumber \\
& = 1 - \exp\left(-\gamma p(v)q(L)\barn \right), \nonumber
\end{align}
where the inequality results because for any $x$, $\max(1-x, 0) \leq e^{-x}$ (here $x=\gamma \barn q(u) p(v)$). For notational convenience, let $\ell(L)$ be $q(L)$ multiplied by a fixed scalar:
\begin{equation}\label{eq:rel_L_q}
\ell(L): 
= \gamma \barn q(L).
\end{equation}
We use $\ell(L)$ as a measure for the relative size of $L$. Using this new notation, equation (\ref{eqn:inter11}) can be written as
\begin{equation}\label{eq:a_v}
\Pr[v\in \Gamma(L)] = \Pr[I_L(v) = 1]  \ge  1 - \exp\left(-p(v)\ell(L)\right).
\end{equation}
This matches our intuition: when a node $v \in V$ is more important with a larger $p(v)$ or the subset $L \subseteq U$ is larger, the chance of $v$ being a neighbor of $L$ increases.

\subsection{Warmup analysis: the balanced and symmetrical case}
\label{sec:warmup}

To better illustrate the idea behind the proof of \pref{lem:lefttorightexpansiondemand}, we first prove a weaker version of \pref{lem:lefttorightexpansiondemand} under the balanced and symmetrical setting. In this special case, we assume that $m = n$, $\Es(u) = \Ed(v) = \frac{1}{n}$ for all $u \in U$ and $v\in V$, and $\{d(v): v \in V\}$ are independent. For ease of illustration, we only prove the result of \pref{lem:lefttorightexpansiondemand} for subsets $L \subseteq U$ with $q(L) = \epsilon$. The proof of this special case demonstrates the high-level idea of the actual proof of \pref{lem:lefttorightexpansiondemand}. However, to extend it to the general unbalanced and asymmetrical case, we need several important ingredients to overcome a few technical difficulties (see \pref{sec:general-case-overview}).

\begin{lemma}[Special case of \pref{lem:lefttorightexpansiondemand}]\label{lem:weak}
	Let us assume  $m = n = \barn$, $\Es(u) = \Ed(v) = \frac{1}{n}$ for all $u\in U$ and $v\in V$, and $\{d(v): v \in V\}$ are independent. For any $L \subseteq U$ with $q(L)= \eps$, we have $d(\Gamma(L))\ge \kappa q(L).$
\end{lemma}

We replace $\frac{\kappa}{\ck} q(L)$ on the RHS of \eqref{eqn:inter13} in \pref{lem:lefttorightexpansiondemand} based on $\kappa q(L)$ in \pref{lem:weak}. This replacement does not change the proof idea, but makes the exposition cleaner. Under the assumption of \pref{lem:weak}, $p(\cdot)$ and $q(\cdot)$ reduce to constant functions, and $q(L)$ and $\ell(L)$ become proportional to the size of the subset $L$, that is,
\[
p(v) = \frac{1}{n},~ ~ q(u) = \frac{1}{n}, ~~ q(L) = \frac{|L|}{n}, ~~\text{and}~~~ \ell(L) =\gamma n q(L)= \gamma |L|.
\]


We fix an $L\subseteq U$ with $q(L) = \epsilon$ and omit the subscript $L$ in $I_L(v)$ for notational simplicity. By definition, we have $\ell(L) =\gamma n q(L)  =\gamma \epsilon n = O_{\kappa}(\ln (1/\eps)\eps n) < n$ for small enough $\eps$.  Therefore, we have $p(v)\ell(L) =\frac{1}{n} \ell(L) < 1$. Using (\ref{eq:a_v}), we can approximate that $\Pr[I(v) = 1]$ by
\begin{align}
 \Pr[I(v) = 1]\geq 1 - \exp\left(-p(v)\ell(L)\right) \geq p(v)\ell(L)/2,
 \label{eqn:Iv-lb}
\end{align}
where the last inequality results because $1-\exp(-x) \geq x/2$ for any $x \in [0,1]$.


To prove \pref{lem:weak}, for each $L$ with $q(L) =\eps $, we prove that $d(\Gamma(L))$ is larger than $q(L)$ w.h.p. (as shown in the next lemma), and then take the union bound over $L \subseteq U$.

\begin{lemma}\label{lem:weak-dL}
	Under the assumption of \pref{lem:weak}, we have
		\begin{equation}
		\Pr[d(\Gamma(L)) < \kappa q(L) ] \le \exp\left(-\frac{\gamma \epsilon n}{16\kappa }\right) . \label{eq:lemweak}
		\end{equation}
\end{lemma}

\proof{Proof of \pref{lem:weak-dL}.}
	Recall equation (\ref{eqn:dgammaL}), where we write $d(\Gamma(L))$ as a sum of independent random variables
	$
	d(\Gamma(L)) = \sum_{v\in V} d(v)I(v).
	$	
	We use the Chernoff bound (see \pref{cor:chernoff} in the e-companion) to prove that $d(\Gamma(L))$ is large w.h.p. We first estimate the mean of $d(\Gamma(L))$ based on
\begin{align}\label{eqn:warm-up-mu-lb}
\mu \triangleq \Exp [d(\Gamma(L))] = \sum_{v\in V} \Exp[d(v)I(v)] \ge \sum_{v\in V} \Ed(v) p(v)\ell(L)/2= \frac{\ell(L)}{2n},
\end{align}
where the inequality results because $d(v)$ and $I(v)$ are independent, and \eqref{eqn:Iv-lb}. The last equality results because  $\Ed(v) = p(v) = \frac{1}{n}$. Consider that $\kappa q(L) = \kappa |L|/n < \gamma |L|/(4n) = \mu/2$ and that for each $v \in V$, $d(v) I(v) \in [0, \kappa\Ed(v)]$ is an independent random variable. Based on the Chernoff bound in  \pref{cor:chernoff}, we have
\begin{align}\label{eqn:warm-up-dGamma-ub}
\Pr[d(\Gamma(L)) < \kappa q(L)]  \leq \Pr[d(\Gamma(L)) < \mu/2] \leq \exp\left(-\frac{(1/2)^2\mu}{2 \kappa\Ed(v)}\right) =  \exp\left(-\frac{\mu n}{8 \kappa }\right).
\end{align}
We now combine \eqref{eqn:warm-up-mu-lb} and \eqref{eqn:warm-up-dGamma-ub}, and obtain
\[
\Pr[d(\Gamma(L)) < \kappa q(L)] \leq \exp\left(-\frac{\ell(L)}{16\kappa}\right) =\exp\left(-\frac{\gamma |L|}{16\kappa}\right) =\exp\left(-\frac{\gamma \epsilon n}{16\kappa }\right). \tag*{\Halmos}
\]
\endproof

Given \pref{lem:weak-dL}, we can take the union bound over all $L$ such that $q(L) =\eps$ (i.e., $|L| =\epsilon n$), and obtain
\begin{multline*}
\Pr\left[\forall L \textrm{ with } |L| = \eps n, d(\Gamma(L))\ge \kappa q(L)\right]
\geq 1 - \sum_{L: |L| = \eps n} \Pr\left[d(\Gamma(L))< \kappa q(L)\right]\\
\ge 1- \sum_{L: |L| = \eps n}\exp\left(-\frac{\gamma \epsilon n}{16\kappa }\right)
=  1-  {n\choose \eps n}	\exp\left(-\frac{\gamma \epsilon n}{16\kappa }\right)
 \ge 1- n^{-\omega(1)},
\end{multline*}
where the last step is based on $\gamma = \cgamma\kappa^3\ln(e\kappa)\ln(\frac{4\kappa}{\epsilon})   \gg 16\kappa \ln\left(\frac{e}{\eps}\right)$. This gives the proof of \pref{lem:weak}.

\subsection{Extensions to the unbalanced and asymmetrical case: analysis overview} \label{sec:general-case-overview}
We now discuss the technical difficulties of generalizing the analysis in the previous subsection to the unbalanced and asymmetrical case, and show how we manage to address these difficulties.

To apply the union bound over all possible sets $L$ in \pref{lem:lefttorightexpansiondemand}, we must prove that for each fixed set, the bad event $d(\Gamma(L)) < \kappa q(L)$ happens with tiny probability. That is, we must generalize \eqref{eq:lemweak} in the warmup analysis, where we apply a Chernoff bound. However, in the warmup analysis, we can directly apply the Chernoff bound because $d(\Gamma(L))$ can be written as the sum of independent random variables with the same mean. In contrast, in the general heterogeneous demand case, the corresponding random variables may have significantly different means and variances.

\label{page:failure}
To illustrate this difficulty, let us consider the following direct approach of generalizing the proof of \pref{lem:weak-dL}. We again obtain a lower bound on $\mu$ similar to \eqref{eqn:warm-up-mu-lb} using \pref{prop:dprelation} and \eqref{eqn:Iv-lb} as follows:
\begin{equation}\label{eqn:mu_lower}
\mu\triangleq \Exp [d(\Gamma(L))]  \ge \frac{\ell(L)}{2}  \sum_{v\in V} \Ed(v) p(v)\geq \frac{\ck \ell(L)}{2}  \sum_{v\in V} \Ed(v)^2 \ge \frac{\ck \ell(L)}{2n}\left(\sum_{v \in V}\Ed(v)\right)^2 = \frac{\ck \ell(L)}{2n},
\end{equation}
where the last inequality uses  Jensen's inequality. { Based on the definition of $\ell(L)$ in \eqref{eq:rel_L_q}, it is easy to check that the lower bound in \eqref{eqn:mu_lower} is greater than the RHS of \eqref{eqn:inter13} by a multiplicative $\Theta(\ln(1/\eps))$ factor}. However, as each random variable $d(v) I(v) \in [0, \kappa \Ed(v)]$, where $\Ed(v)$ can be as large as $1/\poly\log n$ (see item 2 in \pref{assumption:main-1}), a direct application of the Chernoff bound would result $\exp(-\poly\log n)$ in the probability bound, which is far from the desired $\exp(-\Omega(n))$ bound.

It is worth noting that $d(v) I(v)$ have different variances (for different $v \in V$). Instead of the Chernoff bound, one possible attempt is to investigate the variances and apply Bernstein's inequality. (See the statement in \pref{thm:bernstein_ineq} in the e-companion.) To apply Bernstein's inequality, we compute the sum of the variance of $d(v)I(v)$ as follows:
\begin{equation}\label{eqn:sigma_upper}
\sigma^2 \triangleq \sum_{v\in V}\Var\left[d(v)I(v)\right] \leq  \sum_{v\in V}\Var\left[d(v)\right] \E[I(v)]\nonumber
\le  \sum_{v\in V}\left(\kappa \Ed(v)^2\right) \left( p(v)\ell(L) \right).  
\end{equation}
Here, $\Var[\cdot]$ denotes the variance of a random variable. Note that $\E[I(v)]= \Pr(I(v)=1) \leq \sum_{u \in L} r(u,v) \leq p(v) \ell(L)$. 
As $p(v)$ and $\Ed(v)$ can be as large as $1/\poly\log n$, the estimated upper bound on $\sigma^2$ can be as large as $\kappa \ell(L)/\poly\log n$. Thus, unfortunately, Bernstein's inequality still results $\exp(-\poly \log n)$ in the probability bound at best, making the next step union bound over $L$ fail.

The reasons for the two failed attempts stem from the looseness of the Chernoff bound or Bernstein's inequality for such a particular type of random variable $d(v) I(v)$. At a high level, the special property of $d(v) I(v)$ can be summarized as follows: although $d(v) I(v)$ has a large variance, the probability that $d(v) I(v)$ becomes the largest possible value is small. 
However, known concentration inequalities such as Chernoff and Bernstein only characterize random variables via their maximum possible values and variances, and therefore cannot make use of this special property. Using the special property of our random variables, we are able to prove a new probability bound (see \pref{lem:Vs} below) that serves as a generalization of \pref{eq:lemweak}.

We now give a more detailed introduction of how to generalize  \pref{eq:lemweak}. First, we set up a few more notations. Recall that in fixing an $L \subseteq U$, we have  $\Pr[I_L(v) = 1]  \ge  1 - \exp(-p(v)\ell(L))$ from \eqref{eqn:Iv-lb}. Ideally, we want to simplify this via the approximation $1 - \exp(-x) \approx x$. However, this approximation is only true when $x$ is small. To handle the case when $p(v)\ell(L)$ is large, we partition the set of demand nodes $V$ into two sets:
\begin{equation}\label{eq:V_L}
V_L = \{v \in V: p(v) \ell(L) > \cth\}\quad  \textrm{ and}  \quad V_L^c = \{v \in V: p(v) \ell(L) \le \cth\},
\end{equation}
for some constant $\cth = \cthv\ln(e\kappa)$.  
By setting the threshold $\cth$, $V_L$ is the set of ``large'' nodes $v$, where $\Pr[I_L(v) = 1]$ is very close to 1. That is, a node $v\in V_L$ is quite important in putting 
$v$ in the neighborhood of $L$ with a large probability. The complement of $V_L$, denoted by $V_L^c$, is the set of those ``small'' nodes $v$ where  $\Pr[I_L(v) = 1]$ is bounded away from $1$. For example, when $m \leq n$, for a typical node with $\Ed(v) \approx \frac{1}{n}$ and for a small $L$ with $q(L) \approx \eps$, as the importance value $p(v)\approx \frac{1}{n}$,  $\ell(L) = \gamma \barn q(L)  < \barn=n$, according to \eqref{eq:V_L}, the node $v$ should belong to $V_L^c$.

Given the partition of $V$ in place, we rewrite $d(\Gamma(L))$ as 
\begin{align}\label{eq:decomp}
d(\Gamma(L)) &= \underbrace{\sum_{v\in V_L^c}d(v) I(v)}_{Q_L} + \underbrace{\sum_{v\in V_L}d(v)I(v)}_{W_L}. 
\end{align}
The main rationale behind the partition is that the terms $Q_L$ and $W_L$ are bounded from below in different situations. When $\Ed(V_L^c)$ is large (i.e., $\Ed(V_L^c) = \Omega(1)$), there are many small nodes in $V_L^c$. Therefore, $Q_L$ should concentrate to its expectation, and this concentration is the analog of \pref{eq:lemweak}. We can also estimate the expectation $\E[Q_L] = \sum_{v \in V_L^c} \Ed(v) \E[I(v)] =  \sum_{v \in V_L^c }\Omega( \Ed(v) p(v)\ell(L))$. As $p(v)$ is at least $\Omega(1/n)$ according to our thresholded construction (see \eqref{eq:pq-construction}), we further have $\E[Q_L] = \sum_{v \in V_L^c }\Omega(\Ed(v) \ell(L)/n) = \Omega(\Ed(V_L^c) \ell(L)/n) = \Omega(\ell(L)/n)$. Together with the concentration property of $Q_L$, we should be able to prove that $Q_L$ is $\Omega(\ell(L)/n)$ with very high probability. The following lemma, proved in \pref{sec:Vs}, quantitatively characterizes this intuition. As discussed before, this concentration inequality is novel, as Bernstein/Chernoff is not tight enough for our purpose.

\begin{lemma}\label{lem:Vs}
	If $\Ed(V_L^c)\ge \delta/3$, then 
	\begin{equation}
	\Pr\left[Q_L\ge \frac{\cea \ell(L)}{\kappa n} \right] \geq 1 - \exp\left(-\frac{\cf \ell(L)}{\kappa\ln(e\kappa)}\right),
	\end{equation}	
	where the absolute constant $\delta = 1/\deltav$, $\cea = \ceav$, and $\cf = \cfv$.	
\end{lemma}

However, when $\Ed(V_L)$ is larger than $1-\delta/3$ (i.e., $\Ed(V_L^c)=1-\Ed(V_L) <  \delta/3$), we claim that $W_L$ is at least $(1-\delta)$ w.h.p. As defined in \eqref{eq:decomp}, when $v\in V_L$, $I(v)$ is 1 with a probability very close to 1; therefore, one should expect $W_L$ to be very close to $\sum_{v\in V_L}d(v) = d(V_L)$. As it is assumed that $\Ed(V_L)$ is large, there should be enough terms in the summation $\sum_{v\in V_L}d(v) = d(V_L)$, and $d(V_L)$ (and therefore $W_L$) should concentrate around its mean $\sum_{v\in V_L}\Ed(v) = \Ed(V_L)$. Therefore, the term $W_L$ will be at least $1 - \delta$ w.h.p.

Although the preceding argument on lower bounding the term $W_L$ seems reasonable, the analysis presents a significant difficulty. Even when $\Ed(V_L)$ is large, the probability of $W_L$ failing to concentrate to $\Ed(V_L)$ (and therefore being greater than $1 - \delta$) is not as exponentially small as $\exp(-\Omega(\ell(L))$ in \pref{lem:Vs}, but only as small as $1/\poly (n)$. This prevents us from taking the union bound over exponentially many possible $L$s. This problem mainly results because $W_L$s (for different $L$s) share a common pattern of failure: $d(V_L)$ fails to be greater than $(1 - \delta)$. To solve this problem, we break the cause of the event of $W_L$ failing to be greater than $(1 - \delta)$ into the following two events:
\begin{enumerate}
  \item $d(V_L)$ fails to be greater than $(1 - 2\delta/3)$;
  \item  $W_L$ fails to be greater than $(d(V_L) - \delta/3)$.
\end{enumerate}
For each event separately. we will show that the event does not hold for any $L$ with $\Ed(V_L) \geq 1 - \delta/3$ w.h.p.

For the second event, we follow the classical approach to show that the probability of $W_L$ failing to be greater than $(d(V_L) - \delta/3)$ is exponentially small, and apply the union bound over exponentially many $L$s. For the first event, we make the following crucial observation: although there are exponentially many $L$s, the number of different sets for $V_L$ is only on the order of $O(n)$. To see this, let $V_{\theta} = \{v \in V: p(v) > \theta\}$. We then have $V_L = V_{\theta}$ for $\theta = \cth /\ell(L)$. We call the set $\{V_\theta\}$ the \emph{level sets} of $V$. It is clear that there are only $(n+1)$ different level sets, as $|V|=n$. Therefore, there are at most $(n+1)$ different sets for $V_L$. This important observation allows us to bound the probability of the first event by showing that for each fixed $V_L$, the probability that $d(V_L)$ fails to be greater than $(1-2\delta/3)$ is \emph{polynomially small} (i.e., $n^{-\Omega(1)}$ rather than exponentially small) and then applying the union bound over $O(n)$ possible $V_L$s.

We now formally describe our approach. We first define $\bar{I}(v) = 1 - I(v)$ and split $W_L$ into $W_{L,1}$ and $W_{L,2}$ as follows.
\begin{align}\label{eq:decomp-wl}
W_L \triangleq \sum_{v\in V_L}d(v)I(v) = \underbrace{\sum_{v\in V_L}d(v)}_{W_{L,1}} -\underbrace{\sum_{v\in V_L}d(v)\bar{I}(v)}_{W_{L,2}}.
\end{align}
As discussed previously, $W_{L,1} = d(V_L)$ takes only $(n+1)$ different values for all $L$s and does not depend on the choice of $G$ at all. Therefore, we can simply bound the probability that they are larger than $1-2\delta/3$ for all $L$s via a simple concentration inequality and by taking the union bound over $(n + 1)$ events. In particular, we prove the following lemma for $W_{L, 1}$.
\begin{lemma}\label{lem:V_Lsum-simple}
	With high probability $(1-\barn^{-3})$ over the randomness of $d(\cdot)$, for every $L\subseteq U$ such that $\Ed(V_L) = 1-\Ed(V_L^c)\ge 1-\delta/3$, we have
	$W_{L,1}  \ge 1-\frac{2\delta}{3}.$
\end{lemma}

The term $W_{L,2}$ depends on the graph $G$ due to the term $\bar{I}(v)$. For each fixed $L \subseteq U$, the following lemma shows that $W_{L,2}$ is small with a probability exponentially close to 1.
\begin{lemma}\label{lem:V_Lremainder}
	For any fixed $L\subseteq U$,
	\begin{equation}\label{eq:V_Lremainder}
	\Pr[W_{L,2}\le \delta/3] \ge 1- \exp\left(-\frac{\ch\ell(L)}{\kappa^2}\right), 
	\end{equation}
	where $\delta = \frac{1}{\deltav}$ and $\ch = \chv$.
\end{lemma}
With \pref{lem:V_Lremainder}, we are able to apply the union bound to show that $W_{L, 2} \leq \delta/3$ for all possible $L$s w.h.p. Then, together with \pref{lem:V_Lsum-simple}, we are able to show that $W_L \geq 1 -\delta$ for all possible $L$s w.h.p. The detailed proofs of \pref{lem:V_Lsum-simple} and \pref{lem:V_Lremainder} are relegated to \pref{sec:W} in the e-companion.


\subsection{The formal proof of \pref{lem:lefttorightexpansiondemand} and \pref{lem:lefttorightexpansionp}}
We now have the tools to prove \pref{lem:lefttorightexpansiondemand} (in addition to \pref{lem:lefttorightexpansionp}), which gives our main \pref{thm:main} as shown in \pref{sec:graph_exp}. To prove \pref{lem:lefttorightexpansiondemand}, we take the union bound over every $L$ such that $q(L)\ge \ck\eps/\kappa$ (recall that $\ck =  \ckdisp$ is an absolute constant defined in \pref{lem:lefttorightexpansiondemand}), where the failure probability is controlled by \pref{lem:Vs} and \pref{lem:V_Lremainder} according to whether $\bar{d}(V_L^c)\le \delta/3$. To make the proof flow more smoothly, we extract the union bound calculation in the following lemma, which is used in the proof of \pref{lem:lefttorightexpansiondemand} and \pref{lem:lefttorightexpansionp}. (See its proof in \pref{sec:EC_union} in the e-companion.)

\begin{lemma}\label{lem:union_bound}
	For $\alpha \ge \tbb \barn\ln(2/\zeta)$ with $\zeta \ge \frac{\eps}{2\kappa}$ and sufficiently large $m$ with $\frac{m}{\ln \barn}\ge \frac{6}{\zeta}$, we have
	\begin{equation}
	\sum_{L: q(L)\ge \zeta} \exp\left(-\alpha q(L) \right) \le \barn^{-3}.
	\end{equation}
\end{lemma}

\proof{Proof of \pref{lem:lefttorightexpansiondemand}.}
%

 We first classify $L$ according to whether \pref{lem:Vs} or \pref{lem:V_Lsum-simple} and \pref{lem:V_Lremainder} should be used. Let
\begin{equation}\label{eq:L_1}
\mathcal{L}_1 = \{L: \Ed(V_L) \ge 1-\delta/3\}=\{L: \Ed(V_L^c ) \le \delta/3\} \quad \text{and} \quad  \mathcal{L}_2 = \{L: \Ed(V_L^c)> \delta/3\}.
\end{equation}
It is easy to see that
$\mathcal{L}_1$ and $\mathcal{L}_1$ form a partition of the power set of $U$ (denoted by $2^U$), that is, $\mathcal{L}_1 \cap \mathcal{L}_2 = \emptyset$ and $\mathcal{L}_1 \cup \mathcal{L}_2 = 2^U$. In other words, each $L \subseteq U$ belongs to either $\mathcal{L}_1$ or $\mathcal{L}_2$. We also note that $\mathcal{L}_1$ and $\mathcal{L}_2$ are two deterministic sets and that $p(V_L)$ is a deterministic function that depends only on $L$.

For $L\in \mathcal{L}_1$, let $\mathcal{E}_L$ be the event that $W_{L,2} > \delta/3$, and for $L\in \mathcal{L}_2$, with a slight abuse of notation, let $\mathcal{E}_L$ be the event that $Q_L < \frac{\cea \ell(L)}{\kappa n}$, where $\cea=\ceav$ is defined in \pref{lem:Vs}. Let $\mathcal{F}$ be the event where there exists $L\in \mathcal{L}_1$ such that $W_{L,1} < 1-2\delta/3$.


Note that if none of $\mathcal{E}_L$ happens for all $L$s with $\ck\epsilon/\kappa \le q(L)$, and $\mathcal{F}$ does not happen either, then we can conclude that the following event happens:
\begin{equation}
\forall ~ L\subseteq U \; \textrm{ s.t } \ck\epsilon/\kappa \le q(L),\quad d(\Gamma(L)) \ge \min\left\{1-\delta, \frac{\cea \ell(L)}{\kappa n}\right\}. \label{eqn:event}
\end{equation}

This is achieved in the following case study:
\begin{enumerate}
	\item If $L\in \mathcal{L}_1$, given that $\mathcal{E}_L$ does not happen (i.e., $W_{L,2} \leq \delta/3$) and that $\mathcal{F}$ does not happen (i.e., $\forall L\in \mathcal{L}_1$, $W_{L,1} \ge 1-2\delta/3$), we have
	$$d(\Gamma(L))\ge W_{L,1} - W_{L,2} \ge 1-2\delta/3 -\delta/3 \ge 1-\delta.$$
	\item If $L\in \mathcal{L}_2$, then given that $\mathcal{E}_L$ does not happen (i.e., $Q_L \ge \frac{\cea \ell(L)}{\kappa n}$), we have
	$$d(\Gamma(L))\ge Q_L \ge \frac{\cea \ell(L)}{\kappa n}.$$
\end{enumerate}

Next, we bound the probability that none of the events $\mathcal{E}_L$ and $\mathcal{F}$ happens. Based on the union bound, we have
\begin{equation}
\Pr\left[\left(\bigcup_{L: \ck\epsilon/\kappa \le q(L)}\mathcal{E}_L\right) \bigcup \mathcal{F}\right]\le \Pr[\mathcal{F}]  + \sum_{L: \ck\epsilon/\kappa \le q(L)}\Pr[\mathcal{E}_L] \label{eqn:inter2}
\end{equation}

Based on \pref{lem:Vs}, we have the following: for $L\in \mathcal{L}_2$,
$
\Pr[\mathcal{E}_L]\le \exp\left(-\frac{\cf \ell(L)}{\kappa \ln(e\kappa)}\right),
$
where $\cf=\cfv$.
Based on \pref{lem:V_Lremainder}, we have the following: when $L\in \mathcal{L}_1$,
$
\Pr[\mathcal{E}_L]\le \exp\left(-\frac{\ch\ell(L)}{\kappa^2}\right),
$
where $\ch=\chv$. Therefore, for any $L$ with $\ck\epsilon/\kappa\le q(L)$, we have
$
\Pr[\mathcal{E}_L]\le \exp\left(-\frac{\cfh \ell(L)}{\kappa^2 }\right),
$
for some absolute constant $\cfh = \min\{\cf,\ch\} =\cfhv$. Moreover, based on \pref{lem:V_Lsum-simple}, we know that $\Pr[\mathcal{F}]\le \barn^{-3}$.
Therefore,
\begin{equation}
\textrm{ RHS of } (\ref{eqn:inter2}) \le \barn^{-3} + \sum_{L: \ck\epsilon/\kappa\le q(L)} \exp\left(-\frac{\cfh \ell(L)}{\kappa^2}\right);
\end{equation}
thus, it suffices to prove that
\begin{equation}
\sum_{L: \ck\epsilon/\kappa\le q(L)} \exp\left(-\frac{\cfh \ell(L)}{\kappa^2}\right)\le \barn^{-3}.\label{eqn:inter5}
\end{equation}

Note that $\ell(L) = \gamma \barn q(L)$, and $\gamma\ge \cgamma\kappa^2 \ln(\frac{4\kappa}{\epsilon})$ with $\cgamma = \cgammav$. Then,
$\frac{\cfh \ell(L)}{\kappa^2}$ is at least $\alpha q(L)$ with
\begin{equation}\label{eq:alpha_def}
\alpha=\cfh  \cgamma  \barn  \ln\left(\frac{4\kappa}{\epsilon}\right) \ge \tbb \barn\ln(2/\zeta),
\end{equation}
for $\zeta = \ck\epsilon/\kappa \ge \frac{\eps}{2\kappa}$. We can then invoke \pref{lem:union_bound} and obtain
\begin{equation}
\sum_{L: \ck\epsilon/\kappa\le q(L)} \exp\left(-\frac{\cfh \ell(L)}{\kappa^2}\right)\le \barn^{-3}.
\end{equation}
as desired.

Therefore, the event (\ref{eqn:event}) happens with a probability of at least $(1-2 \bar{n}^{-3})$. Note that $\ell(L) = \gamma \bar{n} q(L)$, and $\gamma\ge \cgamma\kappa^2 \ln\left(\frac{4\kappa}{\epsilon}\right)$ with $\cgamma = \cgammav$. We have
$
\frac{\cea \ell(L)}{\kappa n} \ge \cea \cgamma \ln\left(\frac{4\kappa}{\epsilon}\right) \kappa q(L) \geq \frac{\kappa}{\ck}q(L).
$
Therefore, with probability at least $1-2 \barn^{-3}$,
$
d(\Gamma(L))\ge \min\left\{1-\delta, \frac{\kappa}{\ck} q(L)\right\},
$
which completes the proof. $\qed$
\endproof

We now proceed to prove \pref{lem:lefttorightexpansionp}. Similar to $d(\Gamma(L))$, we decompose with $p(\Gamma(L))$ while simply ignoring the counterpart for $Q_L$,
\begin{align}
p(\Gamma(L)) &\ge \underbrace{\sum_{v\in V_L}p(v)I(v)}_{W^p_L} = \underbrace{\sum_{v\in V_L}p(v)}_{W^p_{L,1}} -\underbrace{\sum_{v\in V_L}p(v)\bar{I}(v)}_{W^p_{L,2}}.
\end{align}

Now we prove \pref{lem:lefttorightexpansionp} by lower bounding $W_{L,1}^p$ and upper bounding $W_{L,2}^p$.

\proof{Proof of \pref{lem:lefttorightexpansionp}.}

	We fix an $L \subseteq U$ with $q(L)\ge \tau = \frac{1}{2\kappa}$. 
	Based on \pref{prop:dprelation}, \eqref{eq:pq-construction}, \eqref{eq:rel_L_q}, and noting  $\gamma\ge \cgamma\kappa^2 \ln(\frac{4\kappa}{\epsilon})$ with $\cgamma = \cgammav$, for any $v \in V$, we have
    \[
	p(v)\ell(L) \ge \frac{5}{6} \cdot \frac{1}{5 n}\cdot \gamma \barn q(L) \geq \frac{\gamma q(L)}{6} \geq \frac{\gamma}{12\kappa} \ge 4\ln(e\kappa). 
    \]
	As $p(v)\ell(L) \ge 4\ln(e\kappa)$ for every $v \in V$, based on the definition of $V_L$ in \eqref{eq:V_L}, we have $V_L = V$. 
	Therefore, we have $W_{L,1}^p = 1.$ 	
	We can bound the term $W_{L,2}^p$ from below based on $$\Pr\left[W_{L,2}^p \ge \delta\right] \le \exp\left(-\frac{\cha\ell(L)}{\kappa^2}\right),$$
where $\cha=\chav$. We separate this lower bound into \pref{lem:p_V_Lremainder} in \pref{sec:W_p} in the e-companion and provide its proof.
Taking the union bound over all $L$s we are interested in, we have
	\begin{equation}
	\Pr\left[\exists L\subseteq U \textrm{ s.t. } q(L)\ge \tau, W_{L,2}^p \ge \delta\right]\le \sum_{L: q(L)\ge \tau}\exp\left(-\frac{\cha\ell(L)}{\kappa^2}\right). \label{eqn:inter22}
	\end{equation}

    Let $\alpha =\frac{\cha \gamma \bar{n}}{\kappa^2}$ so that $\frac{\cha\ell(L)}{\kappa^2} = \alpha q(L)$ based on \eqref{eq:rel_L_q}. We have
    \begin{equation}\label{eq:alpha_p}
     \alpha =\frac{\cha \gamma \bar{n}}{\kappa^2} = \frac{\cha \cgamma \kappa^2 \ln\left(\frac{4\kappa}{\epsilon}\right) \barn}{\kappa^2} \geq \tbb  \barn \ln(2/\zeta),
    \end{equation}
    with $\zeta=\ck \epsilon /\kappa$. As $\epsilon<1/3$ (according to the assumption in the theorem statement), we also have $\zeta \leq \frac{1}{2\kappa} = \tau$. We use \pref{lem:union_bound} and obtain
	\begin{equation}\label{eq:union_W_L_2p}
	\sum_{L: q(L)\ge \tau}\exp\left(-\frac{\cha\ell(L)}{\kappa^2}\right) \leq \sum_{L: q(L)\ge \zeta}\exp\left(-\frac{\cha\ell(L)}{\kappa^2}\right) \le \barn^{-3}.
    \end{equation}
	
	In sum, with a probability of at least $1-\barn^{-3}$, for any $L \subseteq U$ with $q(L)\ge \tau$, we have $W_{L,2}^p < \delta$. 
	It follows that under this event, which happens with a probability of at least $1-\barn^{-3}$,
	\[
    p(\Gamma(L))\ge W_{L,1}^p - W_{L,2}^p \ge 1-\delta,
	\]
which completes the proof. $\qed$
\endproof

\section{Conclusions and Future Directions}
\label{sec:conclusion}
In this paper, we investigate the problem of designing process flexibility to fulfill uncertain demand for a wide class of unbalanced and asymmetrical production systems with heterogeneous random supplies and demands. The proposed design is based on a new thresholding probabilistic construction and is asymptotically optimal. The main idea behind the proof is to reduce the $(1-\eps)$ optimality to generalized expansion properties of the random design. To establish the generalized expansion properties of our construction, we develop new concentration results and an economical way to apply the union bound.

The thresholding idea can be a useful guideline for process flexibility design. Indeed, one key message is that the expected demand of a product is not the only criterion when deciding how many plants should have the corresponding production line. To hedge against the demand uncertainty, even if a product's expected demand is not high, it is still better to give a few different plants the capability to produce this product. This principle can be combined with useful design guidelines. For example, \cite{Chou:11} proposed a greedy heuristic for adding new links to an existing design based on the so-called node expansion ratio. It is interesting to explore the benefit of applying a thresholding scheme to the node expansion ratio. In addition, this paper extends the classical analysis framework for establishing expansion property from a random construction to deal with the heterogeneity of supply and demand. We would like to explore the applications of the generalized expansion property developed in this paper. 
{ Finally, one interesting future direction is to consider cost-sensitive production systems, where different edges have different weights, with the goal of finding a sparse graph with minimum total weight while still fulfilling most of the demand. This is a challenging problem that may require the development of new optimization techniques and spectral graph theory.}


\ACKNOWLEDGMENT{The authors are very grateful to three anonymous referees, the associate editor, and the area editor for their detailed and constructive comments that considerably improved the quality of this paper. We would also like to thank Prof. Christopher Thomas Ryan for helping proofread and edit the paper.
}

\bibliographystyle{ormsv080} 
\bibliography{ref} 

\bigskip

\textbf{Xi Chen} is an assistant professor at Department of Information, Operations, and Management Sciences at Stern School of
Business, New York University. His research interests include statistical machine learning, optimization, and applications of
machine learning to data-driven operations management.

\textbf{Tengyu Ma}  is a visiting researcher at Facebook AI Research. His research focuses on algorithm design and machine learning, including topics such as non-convex optimization, representation learning, deep learning, distributed optimization, high-dimensional statistics.

\textbf{Jiawei Zhang}  is a professor of operations management at the Stern School of Business, New York University, and New York
University Shanghai. His research interests include business analytics and optimization.

\textbf{Yuan Zhou}  is an assistant professor at the computer science department of Indiana University at Bloomington. His research
interests include stochastic and combinatorial optimizations and their applications to operations management and machine learning. He is also
interested in and publishes on analysis of mathematical programming, approximation algorithms, and hardness of approximation.

\ECSwitch


\ECHead{Online Appendix to ``Optimal Design of Process Flexibility for General Production Systems"}

\section{Proof of \pref{lem:global_concentration} }
\label{sec:proof_full}
\medskip
\noindent \textsc{\pref{lem:global_concentration}(restated).} {\it  With high probability over $s(\cdot)$ and $d(\cdot)$,
	\begin{equation*}
	1-\epsilon \leq s(U)\leq 1+\epsilon, \quad \textrm{ and } \quad 1-\epsilon \leq d(V)\leq 1+\epsilon.
	\end{equation*}	
}
\proof{Proof.} According to \pref{assumption:main-1}, we have

\[
\sigma^2 \triangleq \sum_{u \in U} \Var[s(u)] \leq \sum_{u \in U} \kappa \Es(u)^2 \leq  \kappa \left( \sum_{u \in U}  \Es(u)\right) \left(\max_{u \in U} \Es(u) \right) = \kappa \cdot 1 \cdot \frac{O(\eps^2)}{\kappa^3 \ln \bar{n}} = O\left(\frac{\eps^2}{\kappa^2 \ln \bar n}\right),
\]
where $\Var[\cdot]$ denotes the variance of a random variable. Moreover, for each $u \in U$,
\[
|s(u) - \Es(u)| \leq \kappa \Es(u) \triangleq M =  O\left(\frac{\eps^2}{\kappa^2 \ln \bar n}\right).
\]
By the Bernstein inequality for negative associated random variables (\pref{thm:bernstein_ineq} and \pref{cor:concentration-NA}), we have
 \begin{multline}\label{eq:global_conncentration}
\Pr\left[\sum_{u \in U} s(u) < \sum_{u \in U} \Es(u) - \epsilon \right] \leq \exp\left(-\frac{\epsilon^2}{2 \sigma^2 + \frac{2}{3} M \epsilon}\right)\\
 = \exp\left(-\frac{\epsilon^2}{O\left(\frac{\eps^2}{\kappa^2 \ln \bar n}\right) + O\left(\frac{\eps^2}{\kappa^2 \ln \bar n}\right) \epsilon}\right)
\leq \exp\left(- \Omega\left(\kappa^2 \ln \bar n\right)  \right) = \bar{n}^{-\Omega(1)}.
\end{multline}
Therefore, we have $s(U) \geq \Es(U) - \epsilon=1-\epsilon$ with probability $1 - \bar{n}^{-\Omega(1)}$.  By symmetry,  we have $s(U) \leq \Es(U) + \epsilon=1 + \epsilon$ with probability $1 - \bar{n}^{-\Omega(1)}$. By the  union bound, with probability $1 - \bar{n}^{-\Omega(1)}$, we have $1-\epsilon \leq s(U)\leq 1+\epsilon$. Similarly, we can also obtain that $1-\epsilon \leq d(V)\leq 1+\epsilon$ w.h.p.
$\qed$
\endproof

{
\section{Proof of the second statement of \pref{thm:main} via \pref{thm:starting_point}}
\label{sec:reduction}
Let us condition on the event that \eqref{eqn:global_concentration} holds, i.e., $1-\epsilon\leq s(U) \leq 1+\eps$ and $1-\eps \leq d(V) \leq 1+\eps$. We have seen that for any fixed $L \subseteq U$ and $K= V \backslash \Gamma(L)$, either \eqref{eqn:slargerd} or \eqref{eqn:dlargers} implies \eqref{eqn:goal}. For notational simplicity, let $\eqref{eqn:global_concentration}$ denote the event in \eqref{eqn:global_concentration}. Further, let \eqref{eqn:goal}, \eqref{eqn:slargerd} and \eqref{eqn:dlargers} denote the event in \eqref{eqn:goal}, \eqref{eqn:slargerd} and \eqref{eqn:dlargers}   holds for any $L \subseteq U$, respectively.  Using the basic properties of conditional probabilities, we have,
\begin{multline*}
\Pr[\mbox{\eqref{eqn:goal}}] \geq \Pr[\mbox{\eqref{eqn:goal}}~| ~  \mbox{\eqref{eqn:global_concentration}}] \cdot \Pr[\mbox{\eqref{eqn:global_concentration}}] \geq \Pr[\mbox{ \eqref{eqn:slargerd} or \eqref{eqn:dlargers} } |~ \mbox{\eqref{eqn:global_concentration}}]  \cdot \Pr[\mbox{\eqref{eqn:global_concentration}}] \\
=  \Pr[\mbox{ \eqref{eqn:slargerd} or \eqref{eqn:dlargers} }] - \Pr[\mbox{ \eqref{eqn:slargerd} or \eqref{eqn:dlargers} } | ~ \mbox{not \eqref{eqn:global_concentration}}]  \cdot \Pr[\mbox{not \eqref{eqn:global_concentration}}] \geq \Pr[\mbox{ \eqref{eqn:slargerd} or \eqref{eqn:dlargers} }] - \Pr[\mbox{not \eqref{eqn:global_concentration}}],
\end{multline*}
where all the probabilities above are over the choice of $G$ and supply and demand functions. By \pref{lem:global_concentration} we have $ \Pr[\mbox{not \eqref{eqn:global_concentration}}] = \bar{n}^{-\Omega(1)}$; and by \pref{thm:starting_point} we have $\Pr[\mbox{ \eqref{eqn:slargerd} or \eqref{eqn:dlargers} }]  = 1 - \bar{n}^{-\Omega(1)}$. Therefore, together with \pref{lem:max-flow-min-cut-goal}, we have
\begin{equation}\label{eq:G1}
\Pr_{G \sim \calG, s(\cdot), d(\cdot)}[\calZ_G(s, d) \geq 1 - 2\epsilon] \geq \Pr_{G \sim \calG, s(\cdot), d(\cdot)}[\mbox{\eqref{eqn:goal}}] \geq 1 - \zeta^2,
\end{equation}
for some $\zeta=\bar{n}^{-\Omega(1)}$. Let us rewrite
\[
\Pr_{G \sim \calG, s(\cdot), d(\cdot)}[\calZ_G(s, d) \geq 1 - 2\epsilon] = \E_{G \in \mathcal{G}} X,
\]
where $X \triangleq \Pr_{s(\cdot), d(\cdot)} [\calZ_G(s,d) \geq 1 - 2\epsilon] \in [0,1]$ is a random variable. From \eqref{eq:G1}, we have $\E_{G \in \calG} (1-X) \leq \zeta^2$. Combining it with the Markov's inequality, we have
\[
  \Pr_{G \sim \calG} \left[X \geq 1-\zeta\right] =   \Pr_{G \sim \calG} \left[1- X \leq \zeta\right] =1 - \Pr[1-X > \zeta] \geq 1- \frac{\E_{G \sim \calG} (1-X)}{\zeta} \geq 1-\zeta,
\]
which completes the proof of \eqref{eq:G_goal} and thus the second statement of \pref{thm:main}. In other words, with high probability, the random graph $G \sim \calG$ achieves $(1 - 2\epsilon)$-optimality w.h.p.

\section{Proof of \pref{thm:starting_point}}
\label{sec:proof_of_starting}

\noindent \textsc{\pref{thm:starting_point} (restated).} {\it  With high probability over the choice of graph $G$ and supply and demand functions, for any subset $L\subseteq U$, either  \eqref{eqn:slargerd} or \eqref{eqn:dlargers} holds.
}

To prove \pref{thm:starting_point} using  \pref{lem:lefttorightexpansiondemand}  and  \pref{lem:lefttorightexpansionp}, we first introduce the following corollaries of \pref{lem:lefttorightexpansiondemand}  and  \pref{lem:lefttorightexpansionp}.

\begin{corollary}\label{cor:lred}
Assume that $\epsilon < 1/3$. With high probability over the choice of $G$ and the supply and demand functions $s(\cdot)$ and $d(\cdot)$,  for any $L\subseteq U$ such that  $\ck \epsilon/\kappa \le q(L)\le \tau$,
	 \begin{equation}
	 d(\Gamma(L)) \ge \frac{\kappa}{\ck} q(L), \label{eqn:lefttorightexpansionlinear}
	 \end{equation}
where $\tau = \frac{1}{2\kappa}$ and $\ck = \ckdisp$ are two constants.
\end{corollary}
\proof{Proof of \pref{cor:lred}.}
Let us condition on the event in \pref{lem:lefttorightexpansiondemand}. We will prove that the desired event in \pref{cor:lred} happens. When $\ck \eps/\kappa \le q(L)\le \tau= \frac{1}{2\kappa}$ (noting that $\ck \eps < \frac{5}{6}\cdot \frac{1}{3} < \frac{1}{2}$), we have that $1 - \delta =\frac{2}{3} > \frac{1}{2\ck} \ge \frac{\kappa}{\ck} q(L)$. Therefore by \pref{lem:lefttorightexpansiondemand}, we conclude that
	\begin{equation}
	d(\Gamma(L))\ge \min\left\{1-\delta, \; \frac{\kappa}{\ck} q(L)\right\}\ge \frac{\kappa}{\ck} q(L),
	\end{equation}
	which is \eqref{eqn:lefttorightexpansionlinear}.
\qed	
\endproof

Since demand and supply are symmetric, \pref{cor:lred} directly implies the following corollary.

\begin{corollary}\label{cor:rled}
		Assume that $\epsilon < 1/3$. With high probability over the choice of $G$ and the choice of the supply and demand functions $s(\cdot)$ and $d(\cdot)$, for any $K\subseteq V$ such that $\ck\epsilon/\kappa \le p(K)\le \tau$,
			\begin{equation}
			s(\Gamma(K)) \ge\frac{\kappa}{\ck}p(K), \label{eqn:righttoleftexpansionlinear}
			\end{equation}
where $\tau = \frac{1}{2\kappa}$ and $\ck = \ckdisp$ are two constants.
\end{corollary}

\proof{Proof of \pref{thm:starting_point}.}

Let us condition on that the events in \pref{lem:lefttorightexpansionp}, \pref{cor:lred}, and \pref{cor:rled}  happens (which will happen w.h.p. by the union bound). 

Now fixing a subset $L\subseteq U$, we consider the following three cases according to $q(L)$.

\begin{enumerate}
\item If $q(L)\le \ck\epsilon/\kappa$, by \pref{assumption:main-1} and \pref{prop:dprelation}, we have
        \begin{eqnarray*}
            s(L)\le \kappa \Es(L)\le \frac{\kappa}{\ck} q(L)\le \epsilon.
        \end{eqnarray*} \
        Therefore, \eqref{eqn:slargerd}  always holds since the right hand side (RHS) of \eqref{eqn:slargerd} is less than or equal to 0.

\item  If $ \ck\epsilon/\kappa< q(L)\le \tau$, where $\tau = \frac{1}{2\kappa}$, then by \eqref{eqn:lefttorightexpansionlinear} of \pref{cor:lred}, \pref{prop:dprelation} and \pref{assumption:main-1}
		$$d(\Gamma(L))\ge \frac{\kappa}{\ck} q(L) \ge \kappa \Es(L)  \ge s(L).$$
		Therefore, we have that \eqref{eqn:slargerd} holds.
		
\item If  $q(L)> \tau$, where $\tau = \frac{1}{2\kappa}$, then by \eqref{eqn:lefttorightexpansionnonlinear} of \pref{lem:lefttorightexpansionp}, we
have that $p(\Gamma(L))\ge 1-\tau$. Let $K = V\setminus \Gamma(L)$. It follows that $p(K) \le \tau$.  Now we discuss the following two subcases according to the value of  $p(K)$.

       \begin{enumerate}
       \item If $p(K)\le \ck\epsilon/\kappa$, by \pref{prop:dprelation}, we have
         \begin{eqnarray*}
             d(K)\le \kappa \Ed(K)\le \frac{\kappa}{\ck} p(K)\le \epsilon,
         \end{eqnarray*}
          and thus (\ref{eqn:dlargers}) holds since the RHS of \eqref{eqn:dlargers} is less than or equal to 0.
          \item if $\ck\epsilon/\kappa\le p(K)\le \tau$, then by \eqref{eqn:righttoleftexpansionlinear} of \pref{cor:rled},  \pref{prop:dprelation} and \pref{assumption:main-1}
    		$$s(\Gamma(K)) \ge \frac{\kappa}{\ck} p(K) \ge \kappa\Ed(K)\ge d(K),$$
	   	and thus (\ref{eqn:dlargers}) holds.
       \end{enumerate}
\end{enumerate}
By the above case study, we have proved that either \eqref{eqn:slargerd} or \eqref{eqn:dlargers}  is true, which completes the proof of \pref{thm:starting_point}.	 $\qed$
			
\endproof
}


\section{Proofs of \pref{lem:Vs}, \pref{lem:V_Lsum-simple}, and \pref{lem:V_Lremainder}}
\label{sec:EC_lem}
\subsection{Lower bounding $Q_L$}\label{sec:Vs}
%
%

In this subsection, we prove \pref{lem:Vs}. Similar to the standard proof of Chernoff bound, we use the exponential moment method. However, the choice of parameters (e.g. the parameter $t$) in the proof is different from that of Chernoff bound. We first restate the lemma statement.

\medskip
\noindent \textsc{\pref{lem:Vs} (restated).} {\it
	If $\Ed(V_L^c)\ge \delta/3$, then 
	\begin{equation*}
	\Pr\left[Q_L\ge \frac{\cea \ell(L)}{\kappa n} \right] \geq 1 - \exp\left(-\frac{\cf \ell(L)}{\kappa \ln(e\kappa)}\right),
	\end{equation*}
	
	where absolute constant $\delta = 1/\deltav$ and $\cea = \ceav$ and $\cf = \cfv$.
	
}
\medskip

\proof{Proof of \pref{lem:Vs}.}

   Recall the definition $V_L^c=\{v: p(v)\ell(L) \leq c_1\}$ in \eqref{eq:V_L} and $a(v)=\Pr[I(v)=1] \geq 1 - \exp(-p(v) \ell(L))$ in \eqref{eq:a_v}.
By invoking \pref{prop:negexpxle1+Cx} 
with $x = -p(v)\ell(L)$, we have that for any $v\in V_L^c$,
		\begin{equation}\label{eq:av_lbd}
		a(v) = \Pr[I(v) = 1]\ge 1  - \exp(-p(v)\ell(L))  \ge \frac{1-e^{-\cth}}{\cth}p(v)\ell(L)  = \frac{\cb p(v)\ell(L)}{\ln(e\kappa)},
		\end{equation}
		where $\cb = \frac{(1-e^{-\cth})\ln(e\kappa)}{\cth} \ge \cbv$.	

Therefore, for each $v \in V$,
\begin{equation}\label{eq:EdvIv}
\Exp[d(v)I(v)] = \Ed(v) a(v) \geq \frac{\cb \Ed(v) p(v)\ell(L)}{\ln(e\kappa)} .
\end{equation}

Now we begin proving the inequality in the lemma statement by
\begin{multline}\label{eq:pflemVs-expmnt}
\Pr\left[Q_L< \frac{\cea \ell(L)}{\kappa n} \right] = \Pr\left[n Q_L< \frac{\cea \ell(L)}{\kappa} \right] = \Pr\left[n \sum_{v \in V_L^c }d(v)I(v) < \frac{\cea \ell(L)}{\kappa} \right]\\
= \Pr\left[\exp\left(tn \sum_{v \in V_L^c }d(v)I(v)\right) > \exp\left(t \cdot \frac{\cea \ell(L)}{\kappa}  \right)\right],
\end{multline}
where $t < 0$ is an constant that will be decided later. By Markov's inequality, we have
\begin{multline}\label{eq:pflemVs-markov}
\eqref{eq:pflemVs-expmnt} < \Exp\exp\left(tn \sum_{v \in V_L^c }d(v)I(v)\right) \exp\left(-t \cdot \frac{\cea \ell(L)}{\kappa}  \right) \\
\leq  \left(\prod_{v \in V_L^c } \Exp\exp\left(tn d(v)I(v)\right) \right) \exp\left(-t \cdot \frac{\cea \ell(L)}{\kappa}  \right) ,
\end{multline}
where the second inequality is because $d(v)I(v)$ are negative associated (\pref{prop:NA-union}) and \pref{prop:NA-exp}.

Now we state the following claim and defer its proof to the end of proof of \pref{lem:Vs}.
\begin{claim}\label{claim:pflemVs}
$\displaystyle{\Exp \exp(tnd(v)I(v)) \leq  1 +  \frac{a(v)}{\kappa} \left(\exp(tn \kappa \Ed(v))  - 1\right)}$.
\end{claim}
By \pref{claim:pflemVs}, we have
\begin{multline} \label{eq:pflemVs-preav}
\eqref{eq:pflemVs-markov} \leq \exp\left(-t \cdot \frac{\cea \ell(L)}{\kappa}  \right) \prod_{v \in V_L^c } \left(1 +  \frac{a(v)}{\kappa} \left(\exp(tn \kappa \Ed(v))  - 1\right)\right)  \\
\leq \exp\left(-t \cdot \frac{\cea \ell(L)}{\kappa}  \right)   \prod_{v \in V_L^c } \exp\left(\frac{a(v)}{\kappa} \left(\exp(tn \kappa \Ed(v))  - 1\right)\right) \\
= \exp\left(-t \cdot \frac{\cea \ell(L)}{\kappa}  \right)  \exp\left( \sum_{v \in V_L^c } \frac{a(v)}{\kappa} \left(\exp(tn \kappa \Ed(v))  - 1\right)\right) ,
\end{multline}
where the second inequality is by $1 + x \leq e^x$.
Now we apply \eqref{eq:av_lbd} (with $t < 0$ in mind) and get
\begin{multline}\label{eq:pflemVs-afterav}
\eqref{eq:pflemVs-preav} \leq\exp\left(-t \cdot \frac{\cea \ell(L)}{\kappa}  \right)  \exp\left( \sum_{v \in V_L^c } \frac{p(v)\ell(L)}{4 \kappa  \ln(e\kappa)} \left(\exp(tn \kappa \Ed(v))  - 1\right)\right) \\
 \leq\exp\left(-t \cdot \frac{\cea \ell(L)}{\kappa}  \right)  \exp\left( \sum_{v \in V_L^c } \frac{5\Ed(v)\ell(L)}{24 \kappa  \ln(e\kappa)} \left(\exp(tn \kappa \Ed(v))  - 1\right)\right) .
\end{multline}
where the last inequality is by \pref{prop:dprelation} and $t < 0$.

Now we apply \pref{prop:jensenxex} to the second exponential form in \eqref{eq:pflemVs-afterav}, and get
\begin{multline}\label{eq:pflemVs-bfsett}
\eqref{eq:pflemVs-afterav} \leq \exp\left(-t \cdot \frac{\cea \ell(L)}{\kappa}  \right)  \exp\left( \frac{5\ell(L)}{24\kappa \ln(e\kappa)} \cdot  \Ed(V_L^c) \left(\exp\left(\frac{ tn\kappa \Ed(V_L^c)}{|V_L^c|}\right)  -1 \right)\right) \\
\leq \exp\left(-t \cdot \frac{\cea \ell(L)}{\kappa}  \right)  \exp\left( \frac{5\ell(L)}{24\kappa \ln(e\kappa)} \cdot  \Ed(V_L^c) \left(\exp\left({ t\kappa \Ed(V_L^c)}\right)  -1 \right)\right)\\
\leq \exp\left(-t \cdot \frac{\cea \ell(L)}{\kappa}  \right)  \exp\left( \frac{5\ell(L)}{24\kappa \ln(e\kappa)} \cdot  \frac{1}{9} \left(\exp\left({ t\kappa \Ed(V_L^c)}\right)  -1 \right)\right)\\
\leq \exp\left(-t \cdot \frac{\cea \ell(L)}{\kappa}  \right)  \exp\left( \frac{5\ell(L)}{216\kappa \ln(e\kappa)}  \left(\exp\left({ t\kappa/9}\right)  -1 \right)\right),
\end{multline}
where the second inequality used $|V_L^c| \leq |V| = n$, and both the third and the fourth inequalities used the assumption $\Ed(V_L^c) \geq \delta / 3 = 1/9$.

Now we pick $t = -\frac{1}{\ln(e\kappa)}$ and summarize all deductions from \eqref{eq:pflemVs-expmnt} to \eqref{eq:pflemVs-bfsett}, and get
\begin{multline}
\Pr\left[Q_L< \frac{\cea \ell(L)}{\kappa n} \right]  \leq \exp\left( \frac{\cea \ell(L)}{\kappa\ln(e\kappa)}  \right)  \exp\left( \frac{5\ell(L)}{216\kappa \ln(e\kappa)}  \left(\exp\left(-\frac{ \kappa}{9\ln(e\kappa)}\right)  -1 \right)\right)\\
\leq \exp\left( \frac{\cea \ell(L)}{\kappa\ln(e\kappa)}  \right)  \exp\left( \frac{5\ell(L)}{216\kappa \ln(e\kappa)}  \left(\exp\left(-\frac{1}{9}\right)  -1 \right)\right) \leq \exp\left(-\frac{\cf \ell(L)}{\kappa\ln(e\kappa)}\right),
\end{multline}
where the second inequality is because  $\frac{\kappa}{\ln(e\kappa)} \geq 1$ when $\kappa \geq 1$, and we choose
\[
\cf = \cfv \leq -\left(\cea + \frac{5}{216} \left(\exp(-1/9) - 1\right)\right) . \tag*{\Halmos}
\]

\endproof

It remains to prove \pref{claim:pflemVs}.
\proof{Proof of \pref{claim:pflemVs}.}
Let us assume that the random variable $X = d(v) I(v)$ belongs to a discrete probability space $(\Omega, x, p)$. In general cases, similar argument can be made.

Observe that $\Exp X = \sum_{\omega \in \Omega} x(\omega) p(\omega) = \Exp[d(v) I(v)] = \Ed(v) a(v)$, $X \in [0, \kappa \Ed(v)]$, and our goal is to prove that
\[
\Exp \exp(tn X) \leq 1 +   \frac{\Exp X}{\kappa \Ed(v)} \left(\exp(tn \kappa \Ed(v))  - 1\right) .
\]
We begin with
\begin{equation}\label{eq:pflemVs-clm1}
\Exp \exp(tn X) = \sum_{\omega \in \Omega} p(\omega) \exp(tn x(\omega)).
\end{equation}
Now since $\exp(tn x)$ is a convex function of $x$,  for each $\omega \in \Omega$, we have
\begin{multline*}
 \exp(tn x(\omega))  \leq \left(1 - \frac{x(\omega)}{\kappa \Ed(v)}\right) \exp(tn \cdot 0) + \frac{x(\omega)}{\kappa \Ed(v)} \cdot \exp(tn \kappa \Ed(v))  = 1 + \frac{x(\omega)}{\kappa \Ed(v)} \left(\exp(tn \kappa \Ed(v))  - 1\right) .
\end{multline*}
We continue with
\[
\eqref{eq:pflemVs-clm1} \leq \sum_{\omega \in \Omega} p(\omega)  \left( 1 + \frac{x(\omega)}{\kappa \Ed(v)} \left(\exp(tn \kappa \Ed(v))  - 1\right)\right) = 1 +  \frac{\Exp X}{\kappa \Ed(v)} \left(\exp(tn \kappa \Ed(v))  - 1\right) . \tag*{\Halmos}
\]
\endproof

\subsection{Bounds for $W_{L,1}$ and $W_{L,2}$}\label{sec:W}

%


We first prove \pref{lem:V_Lsum-simple}, restated as follows.

\medskip
\noindent\textsc{\pref{lem:V_Lsum-simple} (restated).} {\it
	With high probability ($1-\barn^{-3}$) over the randomness of $d(\cdot)$, for every $L\subseteq U$ such that $\Ed(V_L) = 1-\Ed(V_L^c)\ge 1-\delta/3$,  we have
	\begin{equation*}
	W_{L,1}  \ge 1-2\delta/3.
	\end{equation*}
}
To prove  \pref{lem:V_Lsum-simple}, it suffices to prove the following lemma.
\begin{lemma}\label{lem:V_Lsum}
For any real value $z \ge 0$, let $V_z = \{v: p(v)\ge z\}$. With probability at least $(1-\barn^{-3})$ over the randomness of $d(\cdot)$, we h                                                                  ave that for any $z\ge 0$,
\begin{equation}\label{eq:V_Lsum_d_Vz}
d(V_z) = \sum_{v\in V_z} d(v) \ge \Ed(V_z) - \delta/3 .
\end{equation}
\end{lemma}

Note that for each $L \subseteq U$, we have $V_L=\{v: p(v) \ell(L) \geq c_1\}$ as defined in \eqref{eq:V_L}, and therefore  $V_L$ is $V_z$ with $z=c_1/\ell(L)$ in the definition in \pref{lem:V_Lsum}. As a consequence of \pref{lem:V_Lsum}, with high probability at least $(1-\barn^{-3})$ over the randomness $d(\cdot)$, for any $L\subseteq U$ such that $\Ed(V_L^c)\le \delta/3$ (i.e., $\Ed(V_L) = 1-\Ed(V_L^c)\ge 1-\delta/3$),
\begin{equation}\label{eq:V_Lsum_W_L1}
W_{L,1} = \sum_{v\in V_L} d(v) \ge \Ed(V_L) -\delta/3 \ge 1-2\delta/3,
\end{equation}
and this proves \pref{lem:V_Lsum-simple}.

Now we prove \pref{lem:V_Lsum}.
\proof{Proof of \pref{lem:V_Lsum}.}
We prove \pref{lem:V_Lsum} by applying Bernstein Inequality. First observe that $d(V_z) = \sum_{v\in V_z} d(v)$ is the sum of negative associated random variables with the mean $\E(d(V_z))=\Ed(V_z)$ and the sum of the variances
\begin{multline*}
  \sigma^2 = \sum_{v\in V_z} \Var[d(v)] \leq \sum_{v\in V_z} \E\left[(d(v))^2 \right] \leq \sum_{v\in V_z} \E\left[\kappa \Ed(v)  d(v) \right] =\sum_{v\in V_z} \kappa \Ed(v)^2 \\
    \leq  \kappa \left( \max_{v\in V_z} \Ed(v) \right) \cdot \sum_{v\in V_z}d(v)  \le \kappa\max_{v\in V} \Ed(v) = \kappa M,
\end{multline*}
where
\begin{equation}
    M \triangleq \max_{v\in V}\Ed(v)\le \frac{\cmx}{\kappa^3 \ln \barn } .
\end{equation}
by \pref{assumption:main-1}. Also note $|d(v) -\Ed(v)|\le \kappa \Ed(v) \le \kappa M$ (again by \pref{assumption:main-1}). Applying Bernstein inequality, we have that
\begin{equation}\label{eq:Bernstein_V_z}
\Pr\left[d(V_z) \ge \Ed(V_z) - t\right] \ge 1 -\exp\left(-\frac{t^2}{2\kappa M + 2\kappa M t/3}\right).
\end{equation}
Taking $t = \delta/3$ and $\delta = \frac{1}{\deltav \kappa}$,  we deduce from \eqref{eq:Bernstein_V_z} that
\begin{equation}\label{eq:Bernstein_V_z_1}
\Pr\left[d(V_z) \ge \Ed(V_z) - \delta/3\right] \ge 1 -\exp\left(-4\ln \barn\right) = 1 - \barn^{-4}.
\end{equation}

Although there are infinitely many possible values of $z$, there are only at most $n$ different $V_z$'s. This is because of the monotonicity property of $V_z$, i.e., for $z_1 \leq z_2$, $V_{z_1} \supseteq V_{z_2}$. In other words, if we sort $p(v)$ for $v \in V$ in an increasing order, $V_z$ can only consist of consecutive elements from one with the smallest value $p(v)$ to the one with the largest value. This property allows us to take the union bound over all possible values of $z$,
\begin{equation*}
\Pr\left[\forall z\ge 0, d(V_z) \ge \Ed(V_z) - \delta/3\right] \ge  1 - n \cdot \barn^{-4} \geq \barn^{-3},
\end{equation*}
which completes the proof. $\qed$
\endproof

Now we prove \pref{lem:V_Lremainder}, which simply follows the sub-Gaussian property of $W_{L,2}$.

\medskip
\noindent\textsc{\pref{lem:V_Lremainder} (restated).} {\it
	For any fixed $L\subseteq U$,
	\begin{equation*}
	\Pr[W_{L,2}\le \delta/3] \ge 1- \exp\left(-\frac{\ch\ell(L)}{\kappa^2}\right),
	\end{equation*}
	where  $\delta = \frac{1}{\deltav}$ and $\ch = \chv$.
}

\proof{Proof of \pref{lem:V_Lremainder}.}
		To prove \pref{lem:V_Lremainder}, we invoke \pref{prop:NA-union} and \pref{prop:checking_subgaussian} to show that $d(v)\bar{I}(v)$ for all $v \in V_L$ are negative associated sub-Gaussian random variables and then apply sub-Gaussian concentration to obtain \eqref{eq:V_Lremainder}.
Recall $V_L = \{v: p(v) \ell(L) > \cth\}$ in \eqref{eq:V_L}. By \eqref{eq:a_v}, for any $v \in V_L$, we have
\begin{eqnarray}
  \Pr[\bar{I}(v) = 1] = \Pr[I(v) = 0]\le \exp(-p(v)\ell(L))\le \exp(-\cth)\le e^{-2},
\end{eqnarray}
where the last inequality is because $\cth=4\ln(e\kappa)$. 

By \pref{prop:checking_subgaussian}, we know that $d(v)\bar{I}(v)$ are sub-Gaussian random variables with the variance proxy
\begin{eqnarray*}
  \frac{6\kappa^2\Ed(v)^2}{\ln\left(\frac{1}{\exp(-p(v)\ell(L))}\right)} =   \frac{6\kappa^2\Ed(v)^2}{p(v)\ell(L)}  \le \frac{216\kappa^2p(v)}{25 \ell(L)},
\end{eqnarray*}
where the inequality is due to the fact that $p(v)\ge \frac{5}{6} \cdot \Ed(v) $ (\pref{prop:dprelation}). Also, by \pref{prop:NA-union}, we know that $d(v)\bar{I}(v)$ are negative associated.
Therefore, $W_{L,2}=\sum_{v\in V_L}d(v)\bar{I}(v)$ is a sub-Gaussian random variable with variance proxy
\[
 \sum_{v\in V_L} \frac{216\kappa^2 p(v) }{25 \ell(L) }=  \frac{216\kappa^2 }{25 \ell(L) }\sum_{v\in V_L} p(v) \le \frac{216\kappa^2}{25\ell(L)}\le \frac{9\kappa^2}{\ell(L)}.
\]
 Applying the sub-Gaussian tail bound, we obtain that,
\begin{equation}
 			\Pr\left[\sum_{v\in V_L}d(v)\bar{I}(v) \ge \E\left[\sum_{v\in V_L}d(v)\bar{I}(v)\right] + t\right] \le \exp\left(-\frac{t^2\ell(L)}{18\kappa^2}\right). \label{eqn:inter6}
\end{equation}
We also note that
\begin{equation}
		\Exp\left[\sum_{v\in V_L} d(v)\bar{I}(v)\right] \le \sum_{v\in V_L} \exp\left(-p(v)\ell(L)\right) \Ed(v) \le e^{-\cth}\sum_{v\in V_L}\Ed(v) \le e^{-\cth}.\label{eqn:inter3}
\end{equation}

%
%
 		

 Let $t = \frac{1}{9} - e^{-\cthv} \leq \frac{\delta}{3} - e^{-\cth}$.
 By (\ref{eqn:inter3}) and (\ref{eqn:inter6}), we have that 		
 		\begin{multline}
 			\Pr\left[\sum_{v\in V_L}d(v)\bar{I}(v) \ge \delta/3\right]  \le \Pr\left[\sum_{v\in V_L}d(v)\bar{I}(v) \ge \E\left[\sum_{v\in V_L}d(v)\bar{I}(v)\right] + t\right] \\
			\le \exp\left(-\frac{\ch\ell(L)}{\kappa^2}\right),
 		\end{multline} 		
 		where the absolute constant $\ch =\chv \leq \frac{t^2}{18} = \frac{\left(\frac{1}{9} - e^{-\cthv}\right)^2}{18}$. $\qed$
\endproof

\subsection{A lower bound for $W^p_{L,2}$}\label{sec:W_p}
We prove a counterpart of \pref{lem:V_Lremainder} for the term $W_{L,2}^p$, which  is needed for proving \pref{lem:lefttorightexpansionp}.

\begin{lemma}\label{lem:p_V_Lremainder}
	For any fixed $L\subseteq U$,
	\begin{align}
	\Pr\left[W_{L,2}^p \ge \delta\right] \le \exp\left(-\frac{\cha\ell(L)}{\kappa^2}\right),
	\end{align}
	where $\delta = 1/3$ and the absolute constant $\cha = \chav$.
\end{lemma}

\proof{Proof of \pref{lem:p_V_Lremainder}.}
    We use the same technique for proving \pref{lem:V_Lremainder} to prove \pref{lem:p_V_Lremainder}. By \pref{prop:checking_subgaussian} (treating $p(v)$ as a constant random variable), we know that $p(v)\bar{I}(v)$ are independent sub-Gaussian random variables with the variance proxy
    \begin{eqnarray*}
  \frac{6 p(v)^2}{\ln\left(\frac{1}{\exp(-p(v)\ell(L))}\right)}=\frac{6 p(v)}{\ell(L)} .
\end{eqnarray*}
 Therefore,  $W_{L,2}^p=\sum_{v\in V_L}p(v)\bar{I}(v)$ is a sub-Gaussian random variable with variance proxy
     \[
    \sum_{v\in V_L} \frac{6 p(v)}{\ell(L)} \le \frac{6}{\ell(L)}.
    \]
    Applying the sub-Gaussian tail bound, we obtain that,
    	\begin{equation}
	\Pr\left[\sum_{v\in V_L}p(v)\bar{I}(v) \ge \E\left[\sum_{v\in V_L}p(v)\bar{I}(v)\right] + t\right] \le \exp\left(-\frac{t^2\ell(L)}{12}\right). \label{eqn:inter9}
	\end{equation}
    We also note that
	\begin{equation}
	\Exp\left[W_{L,2}^p\right] = \Exp\left[\sum_{v\in V_L} p(v)\bar{I}(v)\right] \le \sum_{v\in V_L} \exp\left(-p(v)\ell(L)\right) p(v) \le e^{-\cth}\sum_{v\in V_L}p(v) \le e^{-\cth},\label{eqn:inter7}
	\end{equation}
       where we used $p(v)\ell(L) > \cth = 4\ln(e\kappa)$ by the definition of $V_L$ in \eqref{eq:V_L}.
	Let $t = \frac{1}{3} - \frac{1}{e^4} \leq \delta - e^{-\cth} = \frac{1}{3} - \exp(-4\ln(e\kappa))$. By (\ref{eqn:inter9}) and (\ref{eqn:inter7}), we have that
	\begin{align}
	\Pr\left[\sum_{v\in V_L}p(v)\bar{I}(v) \ge \tau\right]  \le \Pr\left[\sum_{v\in V_L}p(v)\bar{I}(v) \ge \E\left[\sum_{v\in V_L}p(v)\bar{I}(v)\right] + t\right] \le \exp\left(-\frac{\cha\ell(L)}{\kappa^2}\right),
	\end{align}	
	where absolute constant  $\cha =\chav \leq \frac{t^2}{12} = \frac{\left(\frac{1}{3} - \frac{1}{e^4} \right)^2}{12}$. $\qed$
\endproof

\section{Proof of \pref{lem:union_bound}}
\label{sec:EC_union}

\medskip
\noindent\textsc{\pref{lem:union_bound} (restated).} {\it
For $\alpha \ge \tbb \barn\ln(2/\zeta)$ with $\zeta \ge \frac{\eps}{2\kappa}$ and sufficiently large $m$ with $\frac{m}{\ln \barn}\ge \frac{6}{\zeta}$, we have
	\begin{equation*}
	\sum_{L: q(L)\ge \zeta} \exp\left(-\alpha q(L) \right) \le \barn^{-3}.
	\end{equation*}
}
\proof{Proof of \pref{lem:union_bound}.}
	To prove this, we argue that it suffices to prove  a stronger inequality
	\begin{equation}
	\sum_{L\subseteq U} \exp(q(L)\barn)\cdot \exp(-\alpha q(L))\le \barn^{-3}\cdot \exp\left(\zeta\barn\right)\label{eqn:inter4}
	\end{equation}
	
	To see why \eqref{eqn:inter4} implies the result in the lemma statement,
	\begin{eqnarray*}
	\textrm{ LHS of } \eqref{eqn:inter4} \ge  \sum_{L: q(L)\ge \zeta} \exp(q(L)\barn)\cdot \exp(-\alpha q(L))
	 \ge \exp(\zeta \barn)\sum_{L: q(L)\ge \zeta}  \exp(-\alpha q(L)),
	\end{eqnarray*}
	and it follows from \eqref{eqn:inter4} that
	$$\sum_{L: q(L)\ge \zeta} \exp(-\alpha q(L)) \le \barn^{-3}.$$
	as desired. 
	
	Now we prove \eqref{eqn:inter4}. Using the binomial expansion theorem and the fact that $q(L) = \sum_{u\in L} q(u)$ 
	we can decompose the LHS of \eqref{eqn:inter4} as follows,
	\begin{multline}
	\sum_{L\subseteq U} \exp(q(L)\barn)\cdot \exp(-\alpha q(L))=  \sum_{L\subseteq U} \prod_{u \in L} \exp(q(u)\barn)\cdot \exp(-\alpha q(u))  \\
	=  \prod_{u\in U}\left(1+ \exp(q(u)\barn) \cdot \exp(-\alpha q(u))\right). \label{eqn:inter10}
	\end{multline}

	Therefore, since $\alpha \ge  \tbb \barn\ln(4\kappa/\epsilon) \ge \tb \barn\ln(4\kappa/\epsilon) + \bar{n}$ and  $\zeta \ge \frac{\eps}{2\kappa}$, we have that
	\begin{multline*}
	1+ \exp(q(u)\barn) \cdot \exp(-\alpha q(u))\le 1 + \exp( -\tb \barn\ln(2/\zeta) q(u))\\
	\le 1 + \exp( -\ln(2/\zeta)) = 1+ \zeta/2  \le \exp\left(\zeta/2\right), 
	\end{multline*}
	where the second inequality is by the fact that $q(u)\ge \frac{1}{n_q} \cdot \frac{1}{5m} \geq \frac{1}{6m} \geq \frac{1}{6\bar n}$ since \pref{prop:dprelation} and $\barn\ge m$.
	
	
	Since $m$ is sufficiently large ($\frac{m}{\ln \barn}\ge \frac{6}{\zeta}$), we have that
	\[
	\zeta/2\le \zeta- \frac{3\ln \barn}{m},
    \]
    and therefore, we altogether we have	
	\begin{eqnarray*}
	\textrm{ RHS of } \eqref{eqn:inter10} \le  \prod_{u\in U}\exp\left(\frac{\zeta}{2}\right)
	\le \prod_{u\in U}\exp\left(\zeta - \frac{3\ln \barn}{m}\right)
	 = \barn^{-3}\cdot \exp\left(\zeta m\right)
	 \leq \barn^{-3}\cdot \exp\left(\zeta \barn\right) .
	\end{eqnarray*}
	
	We have proved \eqref{eqn:inter4} and therefore the whole lemma. $\qed$
\endproof

\section{Some Analysis and Probability Tools}
\label{sec:EC_tool}
%
%


\begin{proposition}\label{prop:negexpxle1+Cx}
	For any real $t \le x\le 0$, $e^x \le 1+\frac{e^t-1}{t}\cdot x$.
\end{proposition}

\proof{Proof of \pref{prop:negexpxle1+Cx}.}
	Consider function $f(x) = e^x -1-\frac{e^t-1}{t}\cdot x$, we have that $f(x)$ is convex. Therefore the maximum value is achieved at the boundary. Therefore we have that $f(x)\le \max\{f(0), f(t)\} = 0$ for all $x\in [t,0]$. $\qed$
\endproof

\begin{proposition}\label{prop:jensenxex}
Suppose $x_1, x_2, \dots, x_z \geq 0$. For every $\alpha \leq 0$, we have
\[\displaystyle{\sum_{i = 1}^z x_i \exp(\alpha x_i) \leq \left(\sum_{i  = 1}^z x_i\right) \exp \left(\frac{\alpha}{z} \sum_{i = 1}^z x_i\right)}.\]
\end{proposition}
\proof{Proof of \pref{prop:jensenxex}.}
This proposition can be proved by a straightforward application of Jensen's inequality. $\qed$
\endproof

\begin{theorem}[Chernoff Bound]\label{thm:chernoff}
Suppose $X_1, X_2, \dots, X_n$ are independent random variables taking values in $\{0, 1\}$. Let $X$ denote their sum and let $\mu = \E [X]$ denote the sum's expected value. Then for any $\delta \in (0, 1)$,
\[
\Pr[X \geq (1+\delta)\mu] \leq \exp\left(-\frac{\delta^2\mu}{3}\right) ~~\mbox{and}~~\Pr[X \leq (1-\delta)\mu] \leq \exp\left(-\frac{\delta^2\mu}{2}\right).
\]
\end{theorem}

A simple corollary of \pref{thm:chernoff} is as follows.
\begin{corollary}\label{cor:chernoff}
Suppose $X_1, X_2, \dots, X_n$ are independent random variables taking values in $[0, M]$. Let $X$ denote their sum and let $\mu = \E [X]$ denote the sum's expected value. Then for any $\delta \in (0, 1)$,
\[
\Pr[X \geq (1+\delta)\mu] \leq \exp\left(-\frac{\delta^2\mu}{3M}\right) ~~\mbox{and}~~\Pr[X \leq (1-\delta)\mu] \leq \exp\left(-\frac{\delta^2\mu}{2M}\right).
\]
\end{corollary}

\begin{theorem}[Bernstein's inequality \citep{Bernnett62}]\label{thm:bernstein_ineq}
Let $x_1,\dots, x_n$ be independent variables with finite variance $\sigma_i^2 = \Var[x_i]$ and bounded by $M$ so that $|x_i - \Exp[x_i]|\le M$. Let $\sigma^2 = \sum_i \sigma_i^2$. Then we have
\[\Pr\left[\left|\sum_{i=1}^n x_i - \Exp\left[\sum_{i=1}^n x_i\right] \right| > t\right] \le 2\exp\left(- \frac{t^2}{2\sigma^2 + \frac{2}{3} M t}\right). \]
\end{theorem}

\begin{proposition}\label{prop:checking_subgaussian}
	Suppose random variable $X = I\cdot D$, where $I$ and $D$ are independent random variables such that $I\in \{0, 1\}$, $b = \Pr[I = 1] \le e^{-2}$, and $D$ has the property that $0\le D \le \kappa \E D$ with $\kappa \geq 1$. Let $\Ed = \E D$. Then $X$ is sub-Guassian with the variance proxy $\frac{6\kappa^2\Ed^2}{\ln \frac{1}{b}}$.
\end{proposition}


\proof{Proof of \pref{prop:checking_subgaussian}.}
	We prove that $X$ satisfies the $\psi_2$-condition, i.e. that $\Exp[\exp\left(a^2(X-\E X)^2\right)]\le 2$ for $a^2 =\frac{\ln 1/b}{2\kappa^2\Ed^2}$. Let $\E X = \mu$. We have that $\mu = b \cdot \Ed \le \Ed$, and thus $|D-\mu |\le \kappa \E D = \kappa \Ed$. For $b\le e^{-2}$, we have
	\begin{align*}
	\Exp[\exp\left(a^2(X-\mu)^2\right)] &=  (1-b)\exp(a^2\mu^2) + b\Exp[\exp(a^2(D-\mu)^2))] \\
	&\le (1-b)\exp(a^2 b^2 \bar{d}^2) + b\exp(a^2\kappa^2\Ed^2)\\
	&\le (1-b)\exp\left(\frac{1}{2}b^2\ln\left(\frac{1}{b}\right)\right) + b\exp\left(\frac{1}{2}\ln \left(\frac{1}{b}\right)\right) \\
	&\le 2 .
	\end{align*}
	Therefore, $X$ is a sub-Gaussian random variable with the variance proxy
	\[
	\frac{3}{a^2}=\frac{6\kappa^2\Ed^2}{\ln \frac{1}{b}} \tag*{\Halmos}.
	\]

\endproof

The following property of negative associated random variables can be found in e.g. \cite{JP83}.
\begin{proposition}\label{prop:NA-union}
Let $X_1, X_2, \dots, X_n$ and $Y_1, Y_2, \dots, Y_n$ be two independent sets of negative associated random variables. Then $X_1 Y_1, X_2 Y_2, \dots, X_n Y_n$ are negative associated.
\end{proposition}

\begin{proposition}\label{prop:NA-exp}
Let $X_1, X_2, \dots, X_n$ be negative associated. Then for every real number $\lambda$,
\[
\E \exp \lambda \sum_{i=1}^{n} X_i \leq \prod_{i = 1}^n \E \exp \lambda X_i .
\]
\end{proposition}
\proof{Proof of \pref{prop:NA-exp}.}
When $\lambda \geq 0$, since $\exp \lambda(X_1 + X_2 + \dots + X_{n-1})$ and $\exp \lambda X_n$ are non-decreasing functions, we have
\[
\E \exp \lambda \sum_{i=1}^{n} X_i \leq \left(\E  \exp \lambda  \sum_{i=1}^{n-1} X_i\right) \left(\E \exp \lambda X_n\right) .
\]
By induction on we know that
\[
\E \exp \lambda \sum_{i=1}^{n-1} X_i \leq \prod_{i = 1}^{n-1} \E \exp \lambda X_i .
\]
Therefore
\[
\E \exp \lambda \sum_{i=1}^{n} X_i \leq  \left( \prod_{i = 1}^{n-1} \E \exp \lambda X_i \right) \left(\E \exp \lambda X_n\right) = \prod_{i = 1}^n \E \exp \lambda X_i .
\]
When $\lambda < 0$, by \pref{prop:NA-union} we know that $-X_1, -X_2, \dots, -X_n$ are also negative associated. Therefore
\[
\E \exp \lambda \sum_{i=1}^{n} X_i  = \E \exp (-\lambda) \sum_{i=1}^{n} (-X_i) \leq \prod_{i = 1}^n \E \exp \lambda X_i .
\] $\qed$
\endproof

As a corollary of \pref{prop:NA-exp}, many concentration inequalities for sum of independent random variables also hold for negative associated random variables.
\begin{corollary}\label{cor:concentration-NA}
Chernoff Bound (\pref{thm:chernoff}, \pref{cor:chernoff}) and Bernstein's inequality (\pref{thm:bernstein_ineq}) hold for negative associated random variables.
\end{corollary}
\proof{Proof sketch.}
In the standard proofs of these inequalities (e.g. \cite{AS04, Bernnett62}), the only place that used the independence of random variables is the equality
\[
\E \exp \lambda \sum_{i=1}^{n} X_i = \prod_{i = 1}^n \E \exp \lambda X_i ,
\]
for every real number $\lambda$. One can replace the inequality by the inequality shown in \pref{prop:NA-exp} and the proofs still go through. $\qed$
\endproof

Similarly, we have the following corollary.
\begin{corollary}\label{cor:subgaussian-NA}
Let $X_1, X_2, \dots, X_n$ be negative associated sub-Gaussian random variables with variance proxies $\sigma_1^2, \sigma_2^2, \dots, \sigma_n^2$. Then $X_1 + X_2 + \dots + X_n$ is a sub-Gaussian random variable with variance proxy $\sigma_1^2 + \sigma_2^2 + \dots + \sigma_n^2$.
\end{corollary}


\section{Additional Experiments}
\label{sec:add_exp}

\begin{figure}[!t]
\centering
\subfigure[Density functions]{
  \includegraphics[width=0.45\textwidth, height=5.7cm]{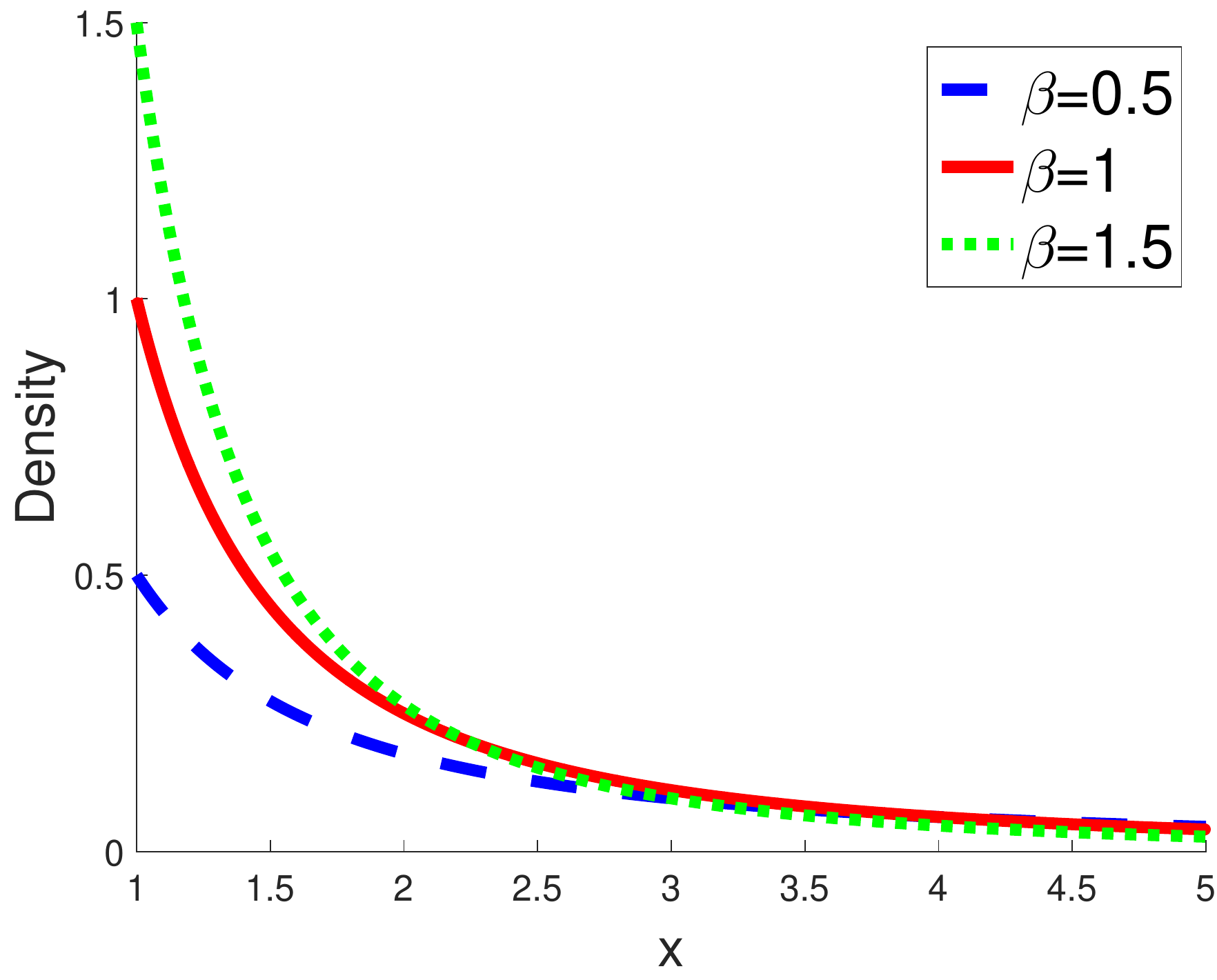}
	    \label{fig:pareto_den}
}
\subfigure[$\beta=0.5$]{
  \includegraphics[width=0.45\textwidth,  height=5.7cm]{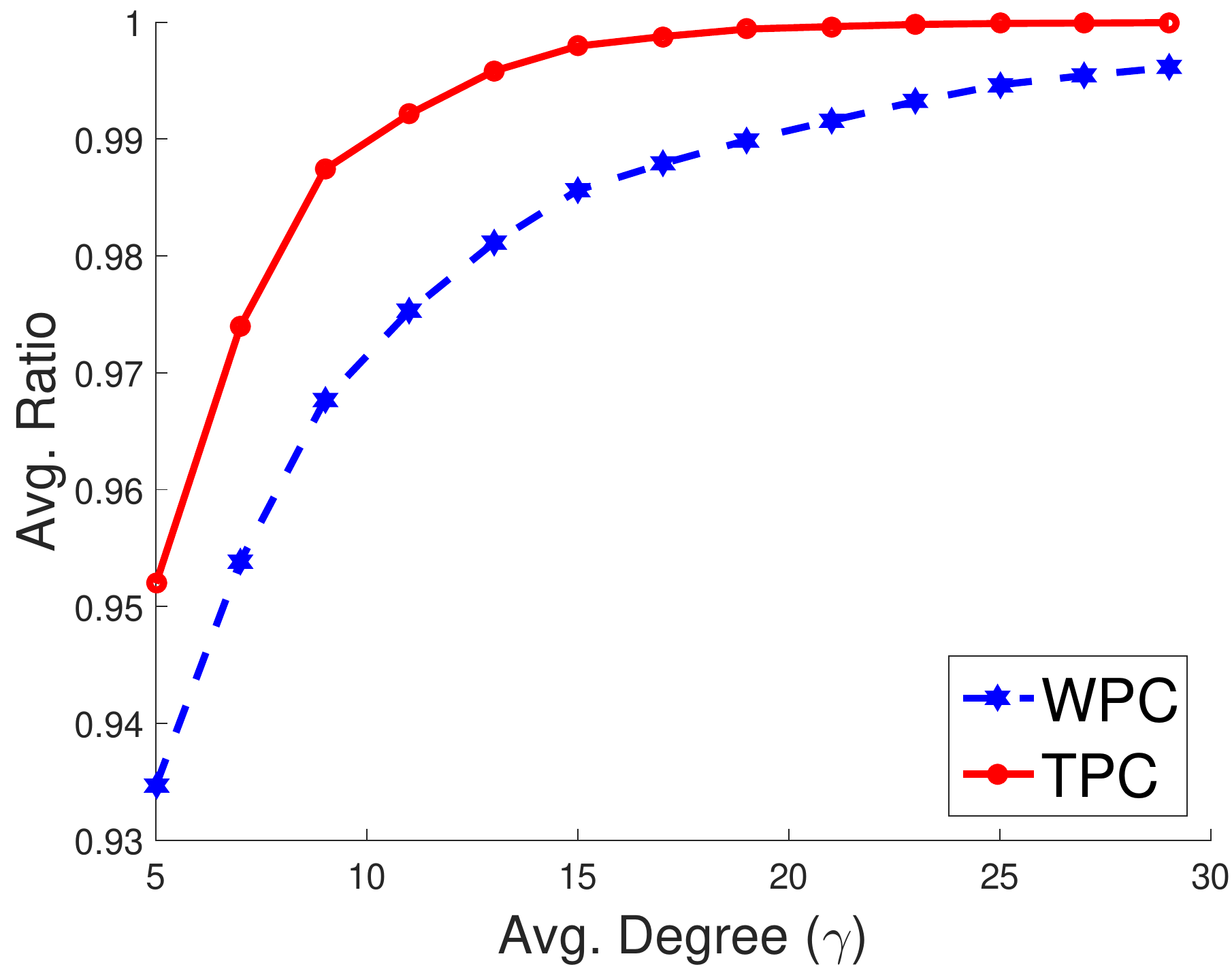}
	    \label{fig:pareto_5}
}
\subfigure[$\beta=1$]{
  \includegraphics[width=0.45\textwidth,  height=5.7cm]{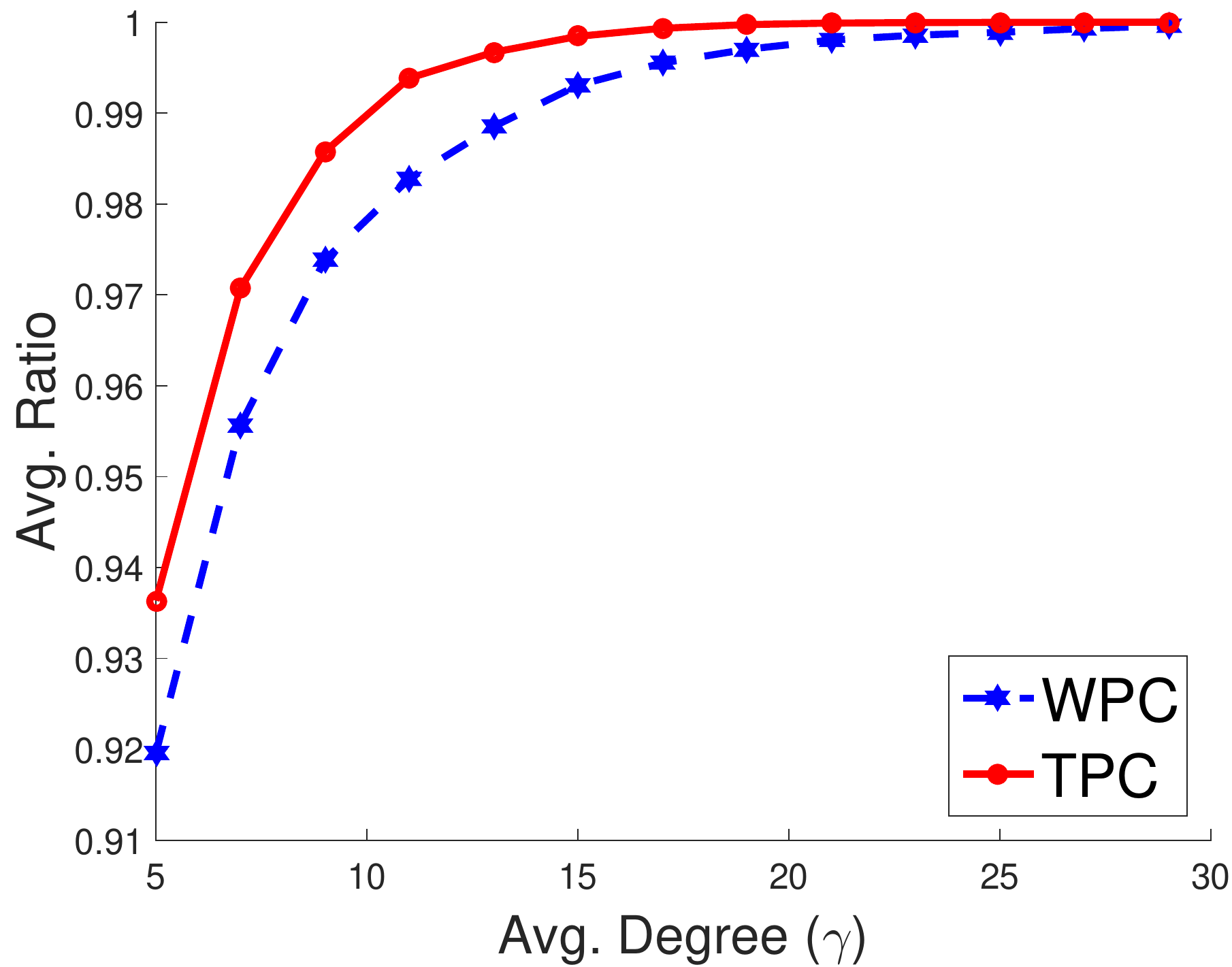}
	    \label{fig:pareto_10}
}
\subfigure[$\beta=1.5$]{
  \includegraphics[width=0.45\textwidth,  height=5.7cm]{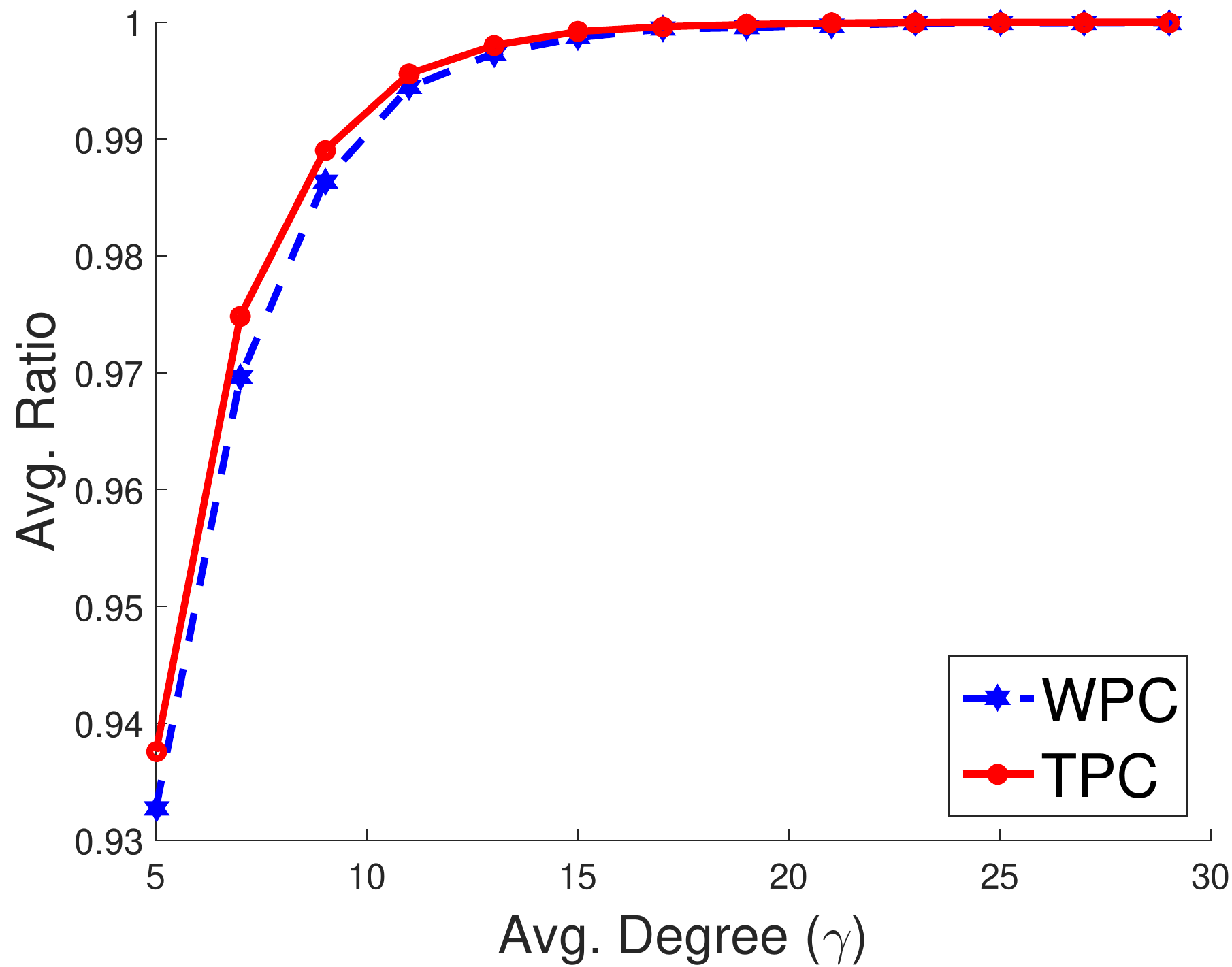}
	    \label{fig:pareto_15}
}
\caption{In (a), we plot the densities of Pareto distributions with different shape parameters $\beta$. In (b)-(d), we present the comparison between the WPC and the TPC when the mean supplies and demands follow the Pareto distribution with different shape parameters (with a truncation at the value 50). The $x$-axis is the average degree $\gamma$, which varies from 5 to 30. The $y$-axis the averaged ratios between the maximum flow of the design $G$ and that of the full flexibility $F$ (the larger the better). Each ratio plotted in the graph is averaged over 100 random graphs for a given $\gamma$ and 1,000 demand realizations. }
\label{fig:comp_pareto}
\end{figure}

In this section, we provide more experiments to compare the effectiveness of the TPC and the WPC. Instead of using the two level supplies as in Section \ref{sec:exp} in the main paper, we study the cases where mean supplies and demands change more smoothly. In particular, we consider the following two setups:
\begin{enumerate}
  \item The mean supplies ($\{\Es(u)\}_{u \in U}$) and mean demands ($\{\Ed(v)\}_{v \in V}$)  are generated from a power law. In particular, we generate the mean supplies and demands from a Pareto distribution with the scale parameter 1 and shape parameter $\beta \in \{0.5,1,1.5\}$ \citep{Newman:04} (see Figure \ref{fig:pareto_den}). We further truncate  excessively large mean supplies and demands to 50, which makes the setting more realistic.
      In fact, the TPC leads to even more significant improvement over the WPC when there is no truncation (since the mean supplies and mean demands  will become more heterogeneous).

  \item The mean supplies ($\{\Es(u)\}_{u \in U}$) and mean demands ($\{\Ed(v)\}_{v \in V}$)  are generated from the uniform distribution on $[0,1]$.
\end{enumerate}

Given the mean supplies ($\{\Es(u)\}_{u \in U}$) (and mean demands ($\{\Ed(v)\}_{v \in V}$), we normalize them so that the sum of the mean supplies (and the mean demands) is 1. We adopt the same experimental setup  as in Section \ref{sec:exp} in the main text. In particular, the $m=100$ supplies are deterministic which take the values of mean supplies, and the $n=100$ demands follow \emph{i.i.d.} two-point distributions.

\begin{figure}[!t]
\centering
\includegraphics[width=0.45\textwidth, height=5.7cm]{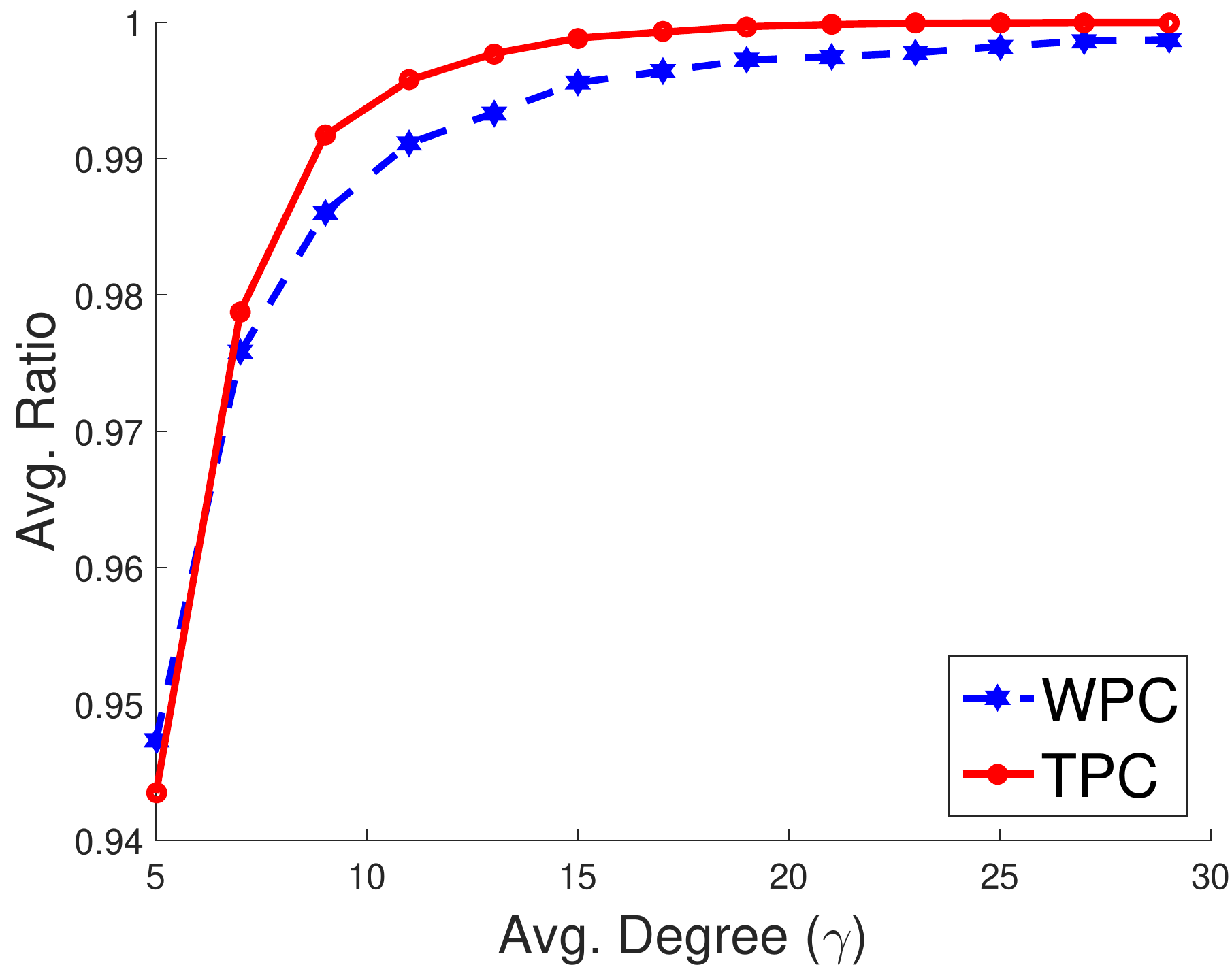}
  \label{fig:uniform}
\caption{We present the comparison between the WPC and the TPC when the mean supplies and demands follow the uniform distribution on $[0,1]$. The $x$-axis is the average degree $\gamma$, which varies from 5 to 30. The $y$-axis the averaged ratios between the maximum flow of the design $G$ and that of the full flexibility $F$ (the larger the better). Each ratio plotted in the graph is averaged over 100 random graphs for a given $\gamma$ and 1,000 demand realizations. }
\label{fig:comp_unif}
\end{figure}

As one can see from Figure \ref{fig:comp_pareto}, the TPC outperforms the WPC when the mean supplies and demands follow a Pareto distribution. When the shape parameter $\beta$ becomes smaller, the corresponding Pareto distribution is more heavy-tailed, which leads to more heterogeneous supplies and demands. In such a case, the improvement of the TPC over the WPC is more significant. The maximum flow of the TPC achieves more than 99\% of the maximum flow of the full flexibility when the average degree $\gamma$ is around 10. The Figure \ref{fig:comp_unif} illustrates the performance comparison between the TPC over the WPC when the mean supplies and demands are drawn from the uniform distribution.






\end{document}